\documentclass[twocolumn, floatfix, prb, aps, showpacs, longbibliography]{revtex4-2}
\usepackage{graphicx, amssymb, color, amsmath}
\usepackage{nicefrac}
\usepackage{multirow, array, booktabs}
\usepackage{float}
\usepackage{romannum}
\usepackage{mathtools}
\usepackage{bbm}
\usepackage{array}
\usepackage{mathrsfs}
\usepackage{IEEEtrantools}
\usepackage{relsize}
\usepackage{mathrsfs}
\usepackage{mathtools}
\usepackage{xparse}
\usepackage[rightcaption]{sidecap}
\usepackage[titletoc, title]{appendix}
\usepackage[colorlinks, bookmarks=true, citecolor=blue, linkcolor=red, urlcolor=blue]{hyperref}
\usepackage{orcidlink}
\usepackage{subcaption}

\graphicspath{{./figures/}}
\begin{document}

\title{Testing the robustness of topological quantities evaluated from the modular Hamiltonian for a given wavefunction}
\author{Sandeep Sharma}
\email{sandeeps@imsc.res.in}
\affiliation{Institute of Mathematical Sciences, CIT Campus, Chennai, 600113, India}
\affiliation{Homi Bhabha National Institute, Training School Complex, Anushakti Nagar, Mumbai 400094, India}

\author{Ajit C. Balram\orcidlink{0000-0002-8087-6015}}
\email{cb.ajit@gmail.com}
\affiliation{Institute of Mathematical Sciences, CIT Campus, Chennai, 600113, India}
\affiliation{Homi Bhabha National Institute, Training School Complex, Anushakti Nagar, Mumbai 400094, India}

\date{\today}

\begin{abstract}
Topologically ordered states are characterized by topological quantities like the Hall conductance, topological entanglement entropy, and chiral central charge. Techniques based on the modular Hamiltonian have recently been developed to extract these quantities from a wavefunction. Here, we consider a lattice model of fractional quantum Hall states, a prototypical example of topologically ordered systems, and extract their topological content using the modular Hamiltonian-based methods. We consider the Laughlin and Moore-Read states and show that the extracted topological quantum numbers converge to their expected results. As expected, the convergence is slower when the correlation length of the state is longer. Generally, our results show that a reliable extraction of topological content through modular methods requires the usage of large systems. 
\end{abstract}

\maketitle
\section{Introduction}
\label{sec: introduction}
Gapped quantum many-body systems in two dimensions (2D) can exhibit phases of matter that carry topological order~\cite{Wen07}. These phases cannot be characterized using any local order parameter and lie beyond the conventional Landau paradigm of classifying phases by the symmetries they spontaneously break~\cite{landau1937, Ginzburg:1950sr}. Topologically ordered matter possesses highly non-local long-range quantum entanglement, hosts Abelian~\cite{Leinaas77, Wilczek82, Arovas84} or non-Abelian~\cite{Moore91, Wen91} excitations that obey anyonic braiding statistics, and has ground state degeneracy~\cite{Wen_1989_Chiral_ground_state_degeneracy, Wen90e} that depends on the topology of the manifold on which the system is placed, and supports chiral gapless modes at its boundary~\cite{Wen90}. These properties are topological in nature, meaning the topological content of the phase is invariant under smooth deformation of the Hamiltonian or the ground state until the bulk energy gap closes, i.e., a quantum critical point is reached, provided the topology of the background manifold on which the system resides remains unchanged.

A paradigmatic example of a topologically ordered phase is a fractional quantum Hall (FQH) fluid~\cite{Tsui82, Laughlin83}. FQH states have a global $U(1)$ symmetry associated with charge conservation, resulting in quantization of the electrical Hall conductance, $\sigma_{xy}{=}(e^2/h)\nu$, where $\nu$ is a rational number called the filling fraction. Another topological quantum number that is a characteristic feature of an FQH phase is its chiral central charge, $c_{-}$, that encodes information on the chiral gapless edge modes that the FQH state supports. The thermal Hall conductance, $\kappa_{xy}$, is related to $c_{-}$ as~\cite{Kane97, Cappelli01a},
\begin{equation}
    \label{eq: Thermal Hall conductance}
    \kappa_{xy}=c_{-}\left(\frac{\pi^2 k_{B}^2}{3h}\right)T
\end{equation}
at a temperature $T$, that is much less than the bulk energy gap, provided the edge modes fully equilibrate. For many FQH states, $\kappa_{xy}$ has been experimentally measured~\cite{Banerjee17, Banerjee17b}, and found to be in good agreement with Eq.~\eqref{eq: Thermal Hall conductance}.

\begin{figure}[t]
    \begin{subfigure}{0.2\textwidth}
        \includegraphics[width=\linewidth]{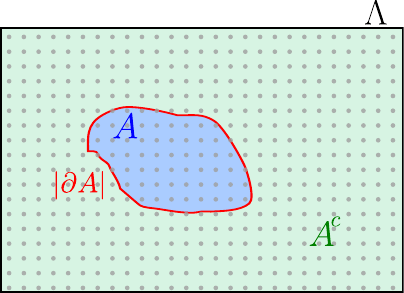}
        \caption{}
        \label{subfig:two_region}
    \end{subfigure}
    \hspace{0.5cm}
    \begin{subfigure}{0.2\textwidth}
        \includegraphics[width=\linewidth]{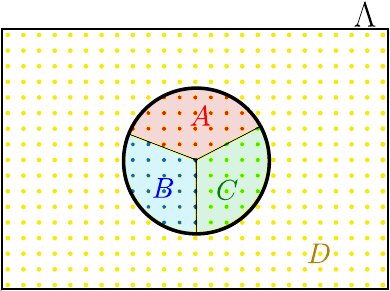}
        \caption{}
        \label{subfig:four_region}
    \end{subfigure}
    \caption{Schematic illustration of a two-dimensional system, $\Lambda$, divided into (a) two regions, $A$ and its complement $A^c$, with boundary $|\partial A|$, and (b) four regions $A$, $B$, $C$, and $D$. The disc-shaped region formed by the union of $A$, $B$, and $C$ in (b) is utilized to extract the topological content of topologically ordered states.
    }
    \label{fig:two_and_four_regions}
\end{figure}

Another topological quantum number that is a feature of topologically ordered phases, albeit one which has not yet been experimentally accessible, is its topological entanglement entropy (TEE)~\cite{Kitaev06, Levin06}, which quantifies its long-range quantum entanglement. Gapped ground states, such as those in the FQH regime, are expected to follow the area law of entanglement entropy. Consider the ground state described by the many-body wavefunction $|\Psi\rangle$. Defining the density matrix $\rho{=}|\Psi\rangle \langle\Psi|$ and partitioning the system into two parts $A$ and its complement $A^{c}$, as shown in Fig.~\ref{subfig:two_region}, the entanglement entropy of the region $A$ is given by
\begin{equation}
    \label{eq: Area law of entanglement}
    S(\rho_A)=\alpha|\partial A|-\gamma +\cdots,
\end{equation}
where $\rho_A$ is the reduced density matrix over the region $A$ that is obtained from tracing out the degrees of freedom in $A^{c}$ on $\rho$, i.e., $\rho_{A}{=}{\rm Tr}_{A^{c}}\left(\rho\right)$. In Eq.~\eqref{eq: Area law of entanglement}, $\alpha$ is a non-universal constant, $|\partial A|$ is the length of the boundary of $A$, $\gamma$ is the TEE, and ellipsis represents terms that vanish in the $|\partial A|{\to} \infty$ limit. As with other topological quantum numbers, the TEE is invariant under smooth deformations of the Hamiltonian, provided a quantum critical point is not reached, and does not depend on the precise shape and size of the region $A$. The TEE is related to the anyonic content of the phase. In particular, for the $\nu{=}1/m$ Laughlin FQH states~\cite{Laughlin83}, that supports $m$ Abelian anyons, the TEE $\gamma_{1/m}^{\rm Laughlin}{=}(1/2)\ln(m)$~\cite{Kitaev06a}.

Recently, a versatile method based on modular Hamiltonians has been devised to extract the above topological quantities from a single bulk wavefunction~\cite{Kim22, Kim22Chiral, Fan23}. For the region $A$, the modular Hamiltonian is defined [in analogy to a thermal state for which $\rho{\propto}e^{-H/(k_B T)}$, where $H$ is the Hamiltonian] as $K_A{=}{-}\ln(\rho_A){\otimes}\mathbb{I}_{A^c}$, where $\mathbb{I}$ is the identity matrix. Topological data is obtained by evaluating the modular Hamiltonian associated with the partitions within the disc-shaped region of Fig.~\ref{subfig:four_region}. Specifically, $c_{-}$ can be obtained from the modular commutator~\cite{Kim22, Kim22Chiral}, defined as the commutator of the modular Hamiltonians over two partially overlapping regions. Similarly, $\sigma_{xy}$ is obtained from the commutator of the modular Hamiltonian and charge operator for two partially overlapping regions~\cite{Fan23}. The TEE is extracted from an appropriate linear combination of the modular Hamiltonians associated with the various regions shown in Fig.~\ref{subfig:four_region}~\cite{Kitaev06, Kim22}.  

In this paper, we test out these methods by numerically evaluating the topological quantum numbers on finite systems for the lattice model of the Abelian Laughlin~\cite{Laughlin83, Nielsen12} and the non-Abelian Moore-Read~\cite{Moore91, Glasser_2015} states. We find that the topological quantum numbers extracted via the modular methods converge to their expected values. As anticipated, the convergence is slower and finite-size effects are stronger when the correlation length of the underlying state is longer. Our results indicate that fairly large systems are necessary to reliably extract topological data using modular techniques.       

The article is organized as follows: Sec.~\ref{sec: models_methods} provides a primer on the background material and is divided into two subsections, Sec.~\ref{ssec: methods} and Sec.~\ref{ssec: candidates}.  
In Sec.~\ref{ssec: methods}, we discuss the modular-Hamiltonian-based techniques we use to compute the various topological quantities of the states. In Sec~\ref{ssec: candidates}, following Refs.~\cite{Nielsen12, Glasser_2015}, we describe the lattice models of the Laughlin and Moore-Read wavefunctions that were derived using correlators of certain conformal operators, for which we will calculate the topological quantities. In Sec.~\ref{sec: results}, we present our numerical results and compare them against theoretical expectations. We conclude the paper in Sec.~\ref{sec: summary} with a summary of our results and provide an outlook for future work.

\section{Methods and model wavefunctions}
\label{sec: models_methods}
In Sec.~\ref{ssec: methods}, we discuss the methodology of extracting topological quantities using the modular Hamiltonian for a given wavefunction. Then, in Sec.~\ref{ssec: candidates}, we discuss the lattice version of two candidate states, namely the Laughlin and Moore-Read states, to which we will apply the modular Hamiltonian methodology to extract their topological content.  

\subsection{Methodology}
\label{ssec: methods}
Consider a 2D many-body gapped system with a locally finite-dimensional Hilbert space, denoted by $\Lambda$, and described by a ground wavefunction $|\Psi\rangle$. The modular Hamiltonian for the subsystem $A$ within $\Lambda$ [see Fig.~\ref{fig:two_and_four_regions}(a)], denoted as $K_A$, is defined as~\cite{Kim22} $K_A{=}{-}\ln\rho_A{\otimes} \mathbb{I}_{A^c}$, where $\rho_A{=}\mathrm{Tr}_{A^c}\!\left(\rho\right)$ is the reduced density matrix of the subsystem $A$ obtained after tracing the degrees of freedom corresponding to the rest of the subsystem, i.e, $A^c$, in the full density matrix, $\rho{=}|\Psi\rangle\langle\Psi|$, of the ground state. To compute the action of $K_A$ on $|\Psi\rangle$, we express $|\Psi\rangle$ as a bipartite state over the regions $A$ and $A^c$. 

Let the Hilbert spaces associated with regions $A$ and $A^c$ be $\mathcal{H}_A$ and $\mathcal{H}_{A^c}$, with basis states $\{|i\rangle_A\}$ and $\{|j\rangle_{A^c}\}$, respectively. The state $|\Psi\rangle$ can be expressed as
\begin{equation}
    \label{eq: bipartite state}
    |\Psi\rangle = \sum_{i,j} m_{ij}\, |i\rangle_A |j\rangle_{A^c}.
\end{equation}
The coefficients $\{m_{ij}\}$ are elements of a rectangular matrix $M$, whose singular value decomposition, $M{=}U D V^\dagger$, results in the Schmidt decomposition of the state $|\Psi\rangle$, i.e., 
\begin{equation}
    \label{eq: bipartite state after Schmidt decomposition}
    |\Psi\rangle = \sum_a d_a\, |a\rangle_A |a\rangle_{A^c},
\end{equation}
where $\{d_a\}$ are the elements of the diagonal matrix $D$. Tracing out the degrees of freedom in region $A^c$, the reduced density matrix for region $A$ for the state $|\Psi\rangle$ is
\begin{equation}
    \label{eq: rho_A after tracing A^c}
        \rho_A = \sum_a d_a^2\, |a\rangle_A\, {}_A\langle a|.
\end{equation}
Then, the action of $K_A$ on $|\Psi\rangle$ is
\begin{equation}
    \label{eq: K_A|state>}
    \begin{split}
        (K_A \otimes \mathbb{I}_{A^c}) |\Psi\rangle = -\sum_a d_a \ln \left(d_a^2\right)\, |a\rangle_A |a\rangle_{A^c}.
    \end{split}
\end{equation}
To calculate the topological quantities of $|\Psi\rangle$, the system is divided into four parts, as shown in Fig.~\ref{subfig:four_region}. The topological invariants are obtained by computing the expectation value of an appropriate combination of the modular Hamiltonian over these regions, as explained below. 

\subsubsection{Hall conductance}
\label{sssec: Hall_conductance}
The Hall conductance is computed by calculating the expectation value of the commutator between the modular Hamiltonian of region $AB$ ($K_{AB}$) and the square of the net charge in region $BC$ ($Q_{BC}^2$)~\cite{Fan23},
\begin{equation}
    \label{eq: Hall Conductance using bulk}
    \sigma_{xy}=\Sigma(\Psi,A,B,C)=\frac{\iota}{2}\langle\Psi|[K_{AB},Q_{BC}^2]|\Psi\rangle,
\end{equation}
where $\iota{=}\sqrt{-1}$ is the imaginary unit. The expression above measures the charge response of the region $BC$ under the modular flow of the region $AB$ and is invariant under smooth deformations of the regions $A$, $B$, and $C$. In units of $e{=}1$, the charge operator $Q_{BC}{=}\sum_{i\in BC} n_i$, where $n_i$ is the occupation number operator of the site $i$ in region $BC$.

\subsubsection{Topological entanglement entropy}
\label{sssec: TEE}
In terms of the modular Hamiltonian, the TEE can be written as
\begin{equation}
    \label{eq: topological entanglement entropy 2}
    \gamma{=}\langle K_{AB} {+} K_{BC} {+} K_{AC} {-} K_A {-} K_B {-} K_C {-} K_{ABC}\rangle_{|\Psi\rangle}.
\end{equation}
As with the Hall conductance, the TEE is also insensitive to smooth deformation of the regions $A$, $B$, and $C$. 

\subsubsection{Chiral central charge}
\label{sssec: chiral_central_charge}
To calculate the chiral central charge from the bulk wavefunction, a quantity called the modular commutator~\cite{Kim22}, denoted as $J(A, B, C)_{|\Psi\rangle}$, is defined over the partition shown in Fig.~\ref{subfig:four_region}, as
\begin{equation}
    \label{eq: modular_comutator}
    J(A,B,C)_{|\Psi\rangle} = i\langle \Psi|[K_{AB},K_{BC}]|\Psi\rangle.
\end{equation}
As with Hall conductance and TEE, $J(A, B, C)_{|\Psi\rangle}$ remains unchanged under smooth deformation of the regions $A$, $B$, and $C$ and is related to the chiral central charge as~\cite{Kim22, Kim22Chiral}:
\begin{equation}
    \label{eq: chiral central charge}
    J(A,B,C)_{|\Psi\rangle}=\frac{\pi}{3}c_{-}.
\end{equation}

\subsection{Candidate wavefunctions}
\label{ssec: candidates}
To apply the aforementioned modular commutator ideas, for $|\Psi\rangle$, we consider two prototypical candidate FQH wavefunctions, namely the Laughlin and Moore-Read states, which we describe in the next two sections.  

\subsubsection{Laughlin}
\label{sssec: Laughlin}
In the continuum planar geometry, the Laughlin wavefunction for $N$ particles is given by~\cite{Laughlin83}:
\begin{equation}
    \label{eq: Laughlin}
    \Psi^{\rm L}_{\nu=1/q} = \prod_{1{\leq}j<k{\leq}N}(z_{j}-z_{k})^{q} \exp\left(-\sum_{l=1}^{N} \frac{|z_l|^{2}}{4\ell^2} \right),
\end{equation}
where $\ell$ is the magnetic length. Positive even integer values of $q$ describe bosonic states, while positive odd integer ones describe fermions. The $q{=}1$ state of Eq.~\eqref{eq: Laughlin} described the Slater determinant $\nu{=}1$ integer quantum Hall (IQH) state. The topological properties of Laughlin states are as follows: 
\begin{itemize}
    \item Hall conductance, $\sigma_{xy}{=}\left( 1/q \right)e^{2}/h$.
    \item topological entanglement entropy, $\gamma{=}\ln(\sqrt{q})$.
    \item chiral central charge, $c_{-}{=}1$.
\end{itemize}
 
Here, we need a lattice version of the Laughlin wavefunction, which was constructed in Ref.~\cite{Nielsen12}. Consider a system of $N$ sites, with $N$ even, in the complex plane, where each site, labeled by the complex numbers $z_{i}$, hosts a spin-$1/2$ particle in the state $s_{i}{=}{\pm }1$. The lattice version of the Laughlin wavefunction in this system is given by~\cite{Nielsen12}
\begin{equation}
    \label{eq: cft Laughlin_lattice s_i}
    |\Psi^{\rm L}_{1,\nu=\frac{1}{4\alpha}} (\{s_i\}_{i=1}^{N})\rangle=\sum_{\{s_i\}_{i=1}^{N}}\psi^{\rm L}_{1,\nu=\frac{1}{4\alpha}} (\{s_i\}_{i=1}^{N})| \{s_i\}_{i=1}^{N} \rangle,
\end{equation}
with the coefficients
\begin{equation}
    \label{eq: cft wfn_coefficients s_i}
    \psi^{\rm L}_{1,\nu=\frac{1}{4\alpha}} (\{s_i\}_{i=1}^{N})=\delta_s\prod_{i=1}^{N}\chi_{i,s_i}\prod_{1{\leq}n<m{\leq}N}(z_n-z_m)^{\alpha s_n s_m},
\end{equation}
where $\delta_s{=}1$ if $\sum_{i{=}1}^N s_i{=}0$, otherwise $\delta_s{=}0$, and $\{\chi_{i,s_i}\}$ are arbitrary phase factors that topological quantities are insensitive to.

It is convenient to write the wavefunction in terms of the occupation number of the sites. The spin state at site $i$ is mapped to the occupation number, $n_i$, as $s_i{=}2n_i{-}1$. In terms of the occupation numbers, the wavefunction of Eq.~\eqref{eq: cft wfn_coefficients s_i} becomes
\begin{equation}
    \label{eq: cft Laughlin_lattice n_i}
    |\Psi^{\rm L}_{2,\nu=\frac{1}{4\alpha}} (\{n_i\}_{i=1}^{N})\rangle=\sum_{\{n_i\}_{i=1}^{N}}\psi^{\rm L}_{2,\nu=\frac{1}{4\alpha}}(\{n_i\}_{i=1}^{N})|\{n_i\}_{i=1}^{N} \rangle,
\end{equation}
with the coefficients
\begin{eqnarray}
    \label{eq: cft wfn_coefficients q_i}
    \psi^{\rm L}_{2,\nu=\frac{1}{4\alpha}}(\{n_i\}_{i=1}^N)&=&\delta_n\prod_{l=1}^{N}\chi_{l,n_l} \prod_{1{\leq}j<k{\leq}N} \Big[(z_j-z_k)^{4\alpha n_j n_k} \nonumber \\
    &&\times  (z_j-z_k)^{-2\alpha (n_j + n_k)}\Big], 
\end{eqnarray}
where $\delta_n{=}1$ if $\sum_{i{=}1}^N n_i{=}N/2$, otherwise $\delta_n{=}0$. As with $\{\chi_{i,s_i}\}$, $\{\chi_{i,n_i}\}$ are arbitrary phase factors that do not affect the topological properties of the underlying phase. In the limit $N{\to}\infty$, we have~\cite{Nielsen12}
\begin{equation}
    \label{eq: Extra factor's large N limit}
    \prod_{1{\leq}j<k{\leq}N} (z_j-z_k)^{-2\alpha (n_j+n_k)} \rightarrow \exp\left( -\frac{\pi \alpha \sum_{i=1}^N n_i|z_i|^2}{b^2} \right),
\end{equation}
up to some phase factor, where $b$ is the lattice spacing. Substituting Eq.~\eqref{eq: Extra factor's large N limit} in Eq.~\eqref{eq: cft wfn_coefficients q_i}, we get
\begin{eqnarray}
     \label{eq: Large N limit wfn_coefficients q_i}
    \psi^{\rm L}_{2,\nu=\frac{1}{4\alpha}}(\{n_i\}_{i=1}^N)&\sim&\delta_n\prod_{i=1}^{N}\chi_{i,n_i} \prod_{1{\leq}j<k{\leq}N} (z_j-z_k)^{4\alpha n_j n_k} \nonumber \\
    &&\times \exp\left({-\frac{\pi \alpha \sum_{i=1}^N n_i|z_i|^2}{b^2}}\right).
\end{eqnarray}
Comparing Eq.~\eqref{eq: Large N limit wfn_coefficients q_i} with Eq.~\eqref{eq: Laughlin}, we conclude that the wavefunction given in Eq.~\eqref{eq: Large N limit wfn_coefficients q_i} is proportional to the Laughlin wavefunction at the filling fraction $\nu{=}(4\alpha)^{-1}$ with the magnetic length, $\ell$, mapping to the lattice spacing $b$, as $\ell{=}b/(2\sqrt{\pi\alpha})$.

Since the topological quantities are determined from the bulk properties of the underlying phase, to avoid the effects of boundaries, we carry out calculations on the compact sphere. The plane to unit sphere mapping is done via the stereographic projection from the south pole, where $z_{i}{=}v_i/u_i$, with $u_i{=}e^{-\iota\phi_i/2} \cos(\theta_i/2)$ and $v_i{=}e^{\iota\phi_i/2}\sin(\theta_i/2)$ being the spinor coordinates corresponding the the polar and azimuthal angles, $\theta_{i}$ and $\phi_{i}$, respectively. On the sphere, the Laughlin wavefunction of Eq.~\eqref{eq: Laughlin} is written as~\cite{Haldane83}:
\begin{equation}
    \label{eq: Laughlin on sphere}
    \Psi_{\nu=1/q}^{\rm L}=\prod_{1{\leq}j<k{\leq}N} (u_j v_k-u_k v_j)^q.
\end{equation}
Substituting this parameterization of $z$ to $(u, v)$ into Eq.~\eqref{eq: cft wfn_coefficients s_i}, we obtain (up to an overall normalization constant)
\begin{equation}
    \label{eq: Laughlin on sphere coefficient 1}
    \psi_{1,\nu=\frac{1}{4\alpha}}^{{\rm L}}(\{s_i\}_{i=1}^N)=\delta_s\prod_{1{\leq}j<k{\leq}N}(u_j v_k-u_k v_j)^{\alpha s_j s_k}.
\end{equation}
Similarly, the Laughlin wavefunction in terms of the occupation numbers is given by:
\begin{equation}
    \label{eq: Laughlin on sphere coefficient 2}
    \psi_{2,\nu=\frac{1}{4\alpha}}^{\rm L} (\{n_i\}_{i=1}^N) = \delta_n\prod_{1{\leq}j<k{\leq}N} (u_j v_k-u_k v_j)^{4\alpha n_j n_k},
\end{equation}
where we have dropped the factor of $\prod_{1{\leq}j<k{\leq}N}  (u_j v_k{-}u_k v_j)^{{-}2\alpha (n_j {+} n_k)}$ from Eq.~\eqref{eq: Large N limit wfn_coefficients q_i} [that turned into the Gaussian factor as $N{\to}\infty$ (see Eq.~\eqref{eq: Extra factor's large N limit})], since it can be shown that the absolute value of the factor $\prod_{1{\leq}j{\leq}N~(j\neq k)}  (u_j v_k{-}u_k v_j)$ does not vary much with $k$ for larger values of $N$, and can be approximated by a constant value for large $N$~\cite{Nielsen12}. We will consider the IQH state at filling $\nu{=}1$ for $\alpha{=}1/4$ and the Laughlin states for $\alpha{=}1/2$, $3/4$, and $1$, corresponding to fillings $\nu{=}1/2$, $1/3$, and $1/4$, respectively, and show results for both the wavefunctions given in Eqs.~\eqref{eq: Laughlin on sphere coefficient 1} and Eqs.~\eqref{eq: Laughlin on sphere coefficient 2}.

\subsubsection{Moore-Read}
\label{sssec: Moore_Read}
In the disk geometry, the Moore-Read wavefunction is written as~\cite{Moore91}:
\begin{equation}
    \label{eq: Moore_Read}
    \Psi^{\rm MR}_{\nu=1/q} = {\rm Pf} \left( \frac{1}{z_{j}-z_{k}} \right) \prod_{1{\leq}j<k{\leq}N}(z_{j}{-}z_{k})^{q} \exp\left(-\sum_{l=1}^{N} \frac{|z_l|^{2}}{4\ell^2} \right),
\end{equation}
where ${\rm Pf}(A)$ is the Pfaffian of the anti-symmetric matrix $A$ with entries $A_{j,k}{=}1/(z_{j}{-}z_{k})$ for $j{\neq}k$ and zero, otherwise. Owing to the anti-symmetry of the Pfaffian, compared to the Laughlin state of Eq.~\eqref{eq: Laughlin}, positive even integer values of $q$ in Eq.~\eqref{eq: Moore_Read} describe fermionic states, while positive odd integer ones describe bosons. The topological properties of the Moore-Read state are as follows: 
\begin{itemize}
    \item Hall conductance, $\sigma_{xy}{=}\left( 1/q \right)e^{2}/h$.
    \item topological entanglement entropy, $\gamma{=}\ln(\sqrt{4q})$.
    \item chiral central charge, $c_{-}{=}3/2$.
\end{itemize}
A lattice model of the Moore-Read state was proposed in Ref.~\cite{Glasser_2015}. In terms of the occupation numbers, this wavefunction is
\begin{equation}
    \label{eq: cft Moore Read n_i}
    |\Psi^{\rm MR}_{\nu=1/q} (\{n_i\}_{i=1}^N)\rangle=\sum_{\{n_i\}_{i=1}^N}\psi^{\rm MR}_{\nu=1/q}(\{n_i\}_{i=1}^N)|\{n_i\}_{i=1}^N \rangle
\end{equation}
with the coefficients
\begin{eqnarray}
    \label{eq: cft Moore Read wfn_coefficients n_i}
    \psi^{\rm MR}_{\nu=1/q}(\{n_i\}_{i=1}^N)&=&\delta_n\prod_{1{\leq}j<k{\leq}N}(z_j-z_k)^{q n_i n_j} \\
    &&\times {\rm Pf}_{n_i=n_j=1}\left(\frac{1}{z_i-z_j}\right)\prod_{l=1}^N f_N (z_l)^{n_l}, \nonumber
\end{eqnarray}
where $f_N (z_l){=}\xi_l\prod_{1{\leq}j{\leq}N~(j\neq l)}(z_j{-}z_l)^{-\eta}$, with $\eta{=}\mathscr{A}/(2\pi)$ with $\mathscr{A}$ being the area per site, $\xi_l$ is a gauge factor, and $\delta_n{=}1$ if $\sum_i n_i{=}\eta N/q$, and $0$ otherwise. When $\eta{=}1$, we get $\sum_i n_i/N{=}1/q$, which is the lattice filling fraction. The continuum limit is obtained as $\eta{\to} 0$ and $N{\to}\infty$. The spherical version of the Moore-Read state is
\begin{eqnarray}
   \label{eq: Moore Read state wfn coefficient on sphere}
    \psi^{\rm MR}_{\nu=1/q}(\{n_i\}_{i=1}^N)&=&\delta_n\prod_{1{\leq}i<j{\leq}N} (u_i v_j - u_j v_i)^{q n_i n_j} \nonumber \\
    && \times {\rm Pf}_{n_i=n_j=1}\left(\frac{1}{u_i v_j - u_j v_i}\right) \\
    && \times \prod_{1{\leq}i<j{\leq}N} (u_i v_j - u_j v_i)^{-\eta (n_i+n_j)}.  \nonumber
\end{eqnarray}
We consider this state at $q{=}1$, $2$, and $3$, corresponding to filling $\nu{=}1$, 1/2 and $1/3$, respectively. We set $\eta{=}1/2$, $1$, and $1$ to study the Moore-Read states at fillings $\nu{=}1$, $1/2$, and $1/3$, respectively. 

\section{Results}
\label{sec: results}
In this section, we present results obtained by numerically calculating the three topological quantities, i.e. the topological entanglement entropy, chiral central charge, and Hall conductance for the lattice model of the Laughlin states given by Eqs.~\eqref{eq: Laughlin on sphere coefficient 1} and~\eqref{eq: Laughlin on sphere coefficient 2} and the Moore-Read state given in Eq.~\eqref{eq: Moore Read state wfn coefficient on sphere}.

\begin{figure}[t]
    \includegraphics[width=0.5\columnwidth]{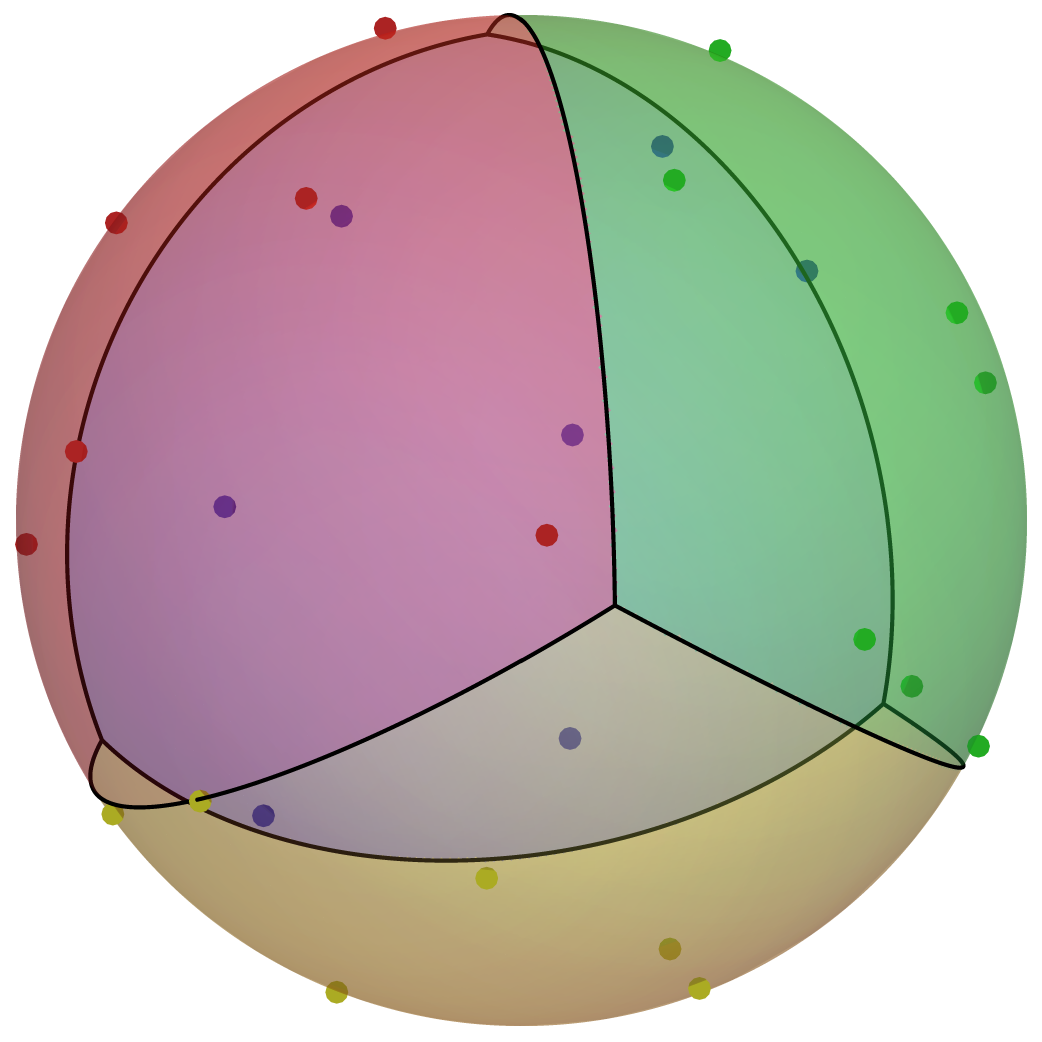}
    \caption{Tetrahedrally symmetric partition of the sphere into four equal parts. The black curves denote the partition boundaries on the sphere, while the colored dots indicate the lattice points on the
    sphere with coordinates described in Eq.~\eqref{eq: coordinate on sphere} for $N{=}26$ sites.
    }
    \label{fig:sphere}
\end{figure}

\begin{figure*}[t]
    \begin{subfigure}{0.12\textwidth}
        {\footnotesize\textbf{state}}
    \end{subfigure}
    \begin{subfigure}{0.28\textwidth}
        \centering
        {\footnotesize \textbf{Hall conductance, $\sigma_{xy}$}}
    \end{subfigure}
    \begin{subfigure}{0.28\textwidth}
        {\footnotesize\textbf{topological entanglement entropy, $\gamma$}}
    \end{subfigure}
    \begin{subfigure}{0.28\textwidth}
        {\footnotesize\textbf{chiral central charge, $c_{-}$}}
    \end{subfigure}

    \vspace{0.3cm}

    \begin{minipage}[c]{0.12\textwidth}
    \begin{subfigure}[b]{1\textwidth}
        {\textbf{$\nu{=}1$ integer quantum Hall}}
    \end{subfigure}
    \end{minipage}
    \begin{minipage}[c]{0.28\textwidth}
        \begin{subfigure}{1\textwidth}
        \includegraphics[width=\linewidth]{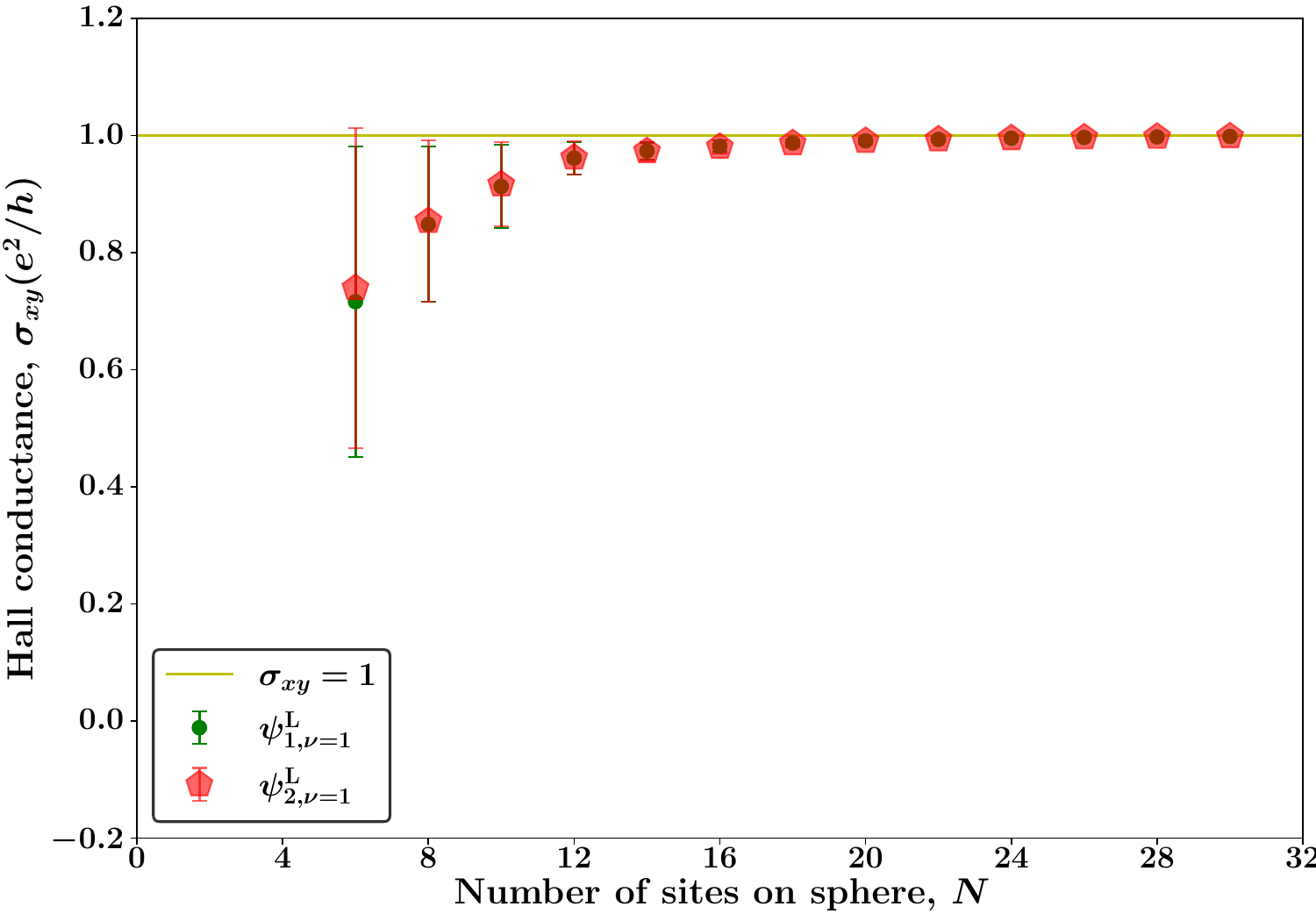}
        \caption{}
        \label{subfig:Hall_1_IQHE}
    \end{subfigure}
    \end{minipage}
    \begin{minipage}[c]{0.28\textwidth}
    \begin{subfigure}{1\textwidth}
        \includegraphics[width=\linewidth]{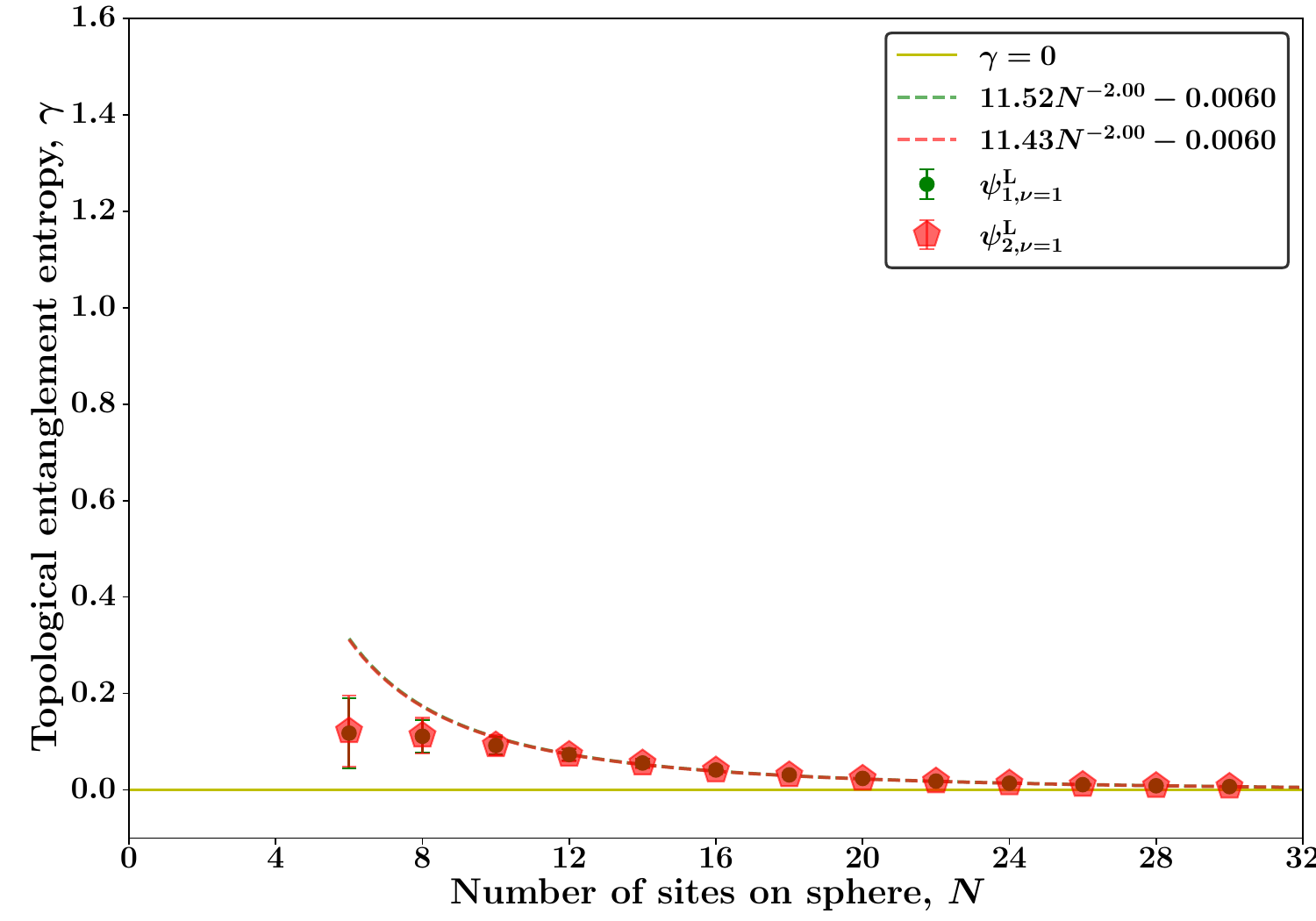}
        \caption{}
        \label{subfig:TEE_1_IQHE}
    \end{subfigure}
    \end{minipage}
    \begin{minipage}[c]{0.28\textwidth}
    \begin{subfigure}{1\textwidth}
        \includegraphics[width=\linewidth]{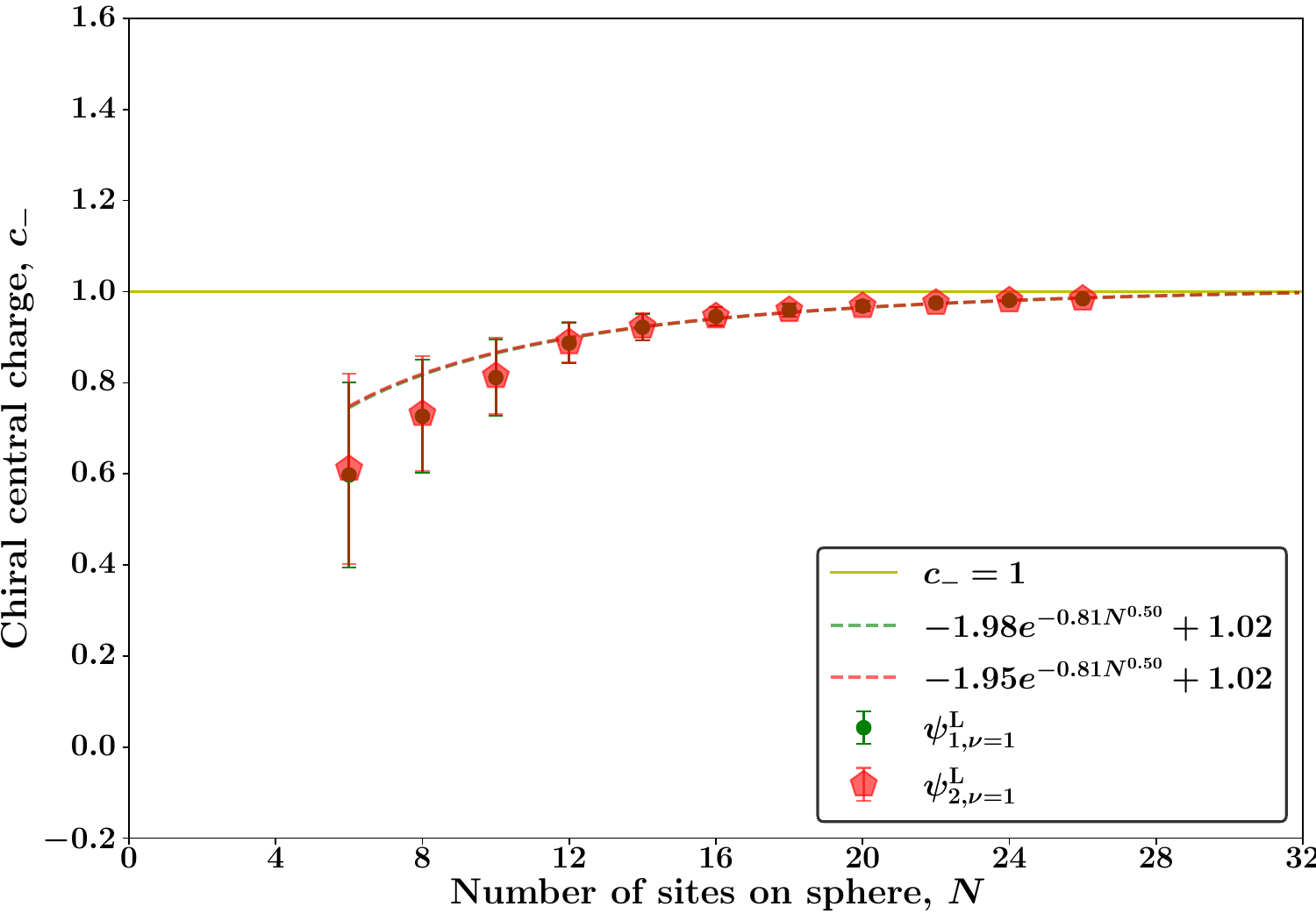}
        \caption{}
        \label{subfig:chiral_central_charge_1_IQHE}
    \end{subfigure}
    \end{minipage}
    \vspace{0.25cm}
    
        \begin{minipage}[c]{0.12\textwidth}
    \begin{subfigure}[b]{1\textwidth}
        {\textbf{$\nu{=}1/2$ Laughlin}}
    \end{subfigure}
    \end{minipage}
    \begin{minipage}[c]{0.28\textwidth}
        \begin{subfigure}{1\textwidth}
        \includegraphics[width=\linewidth]{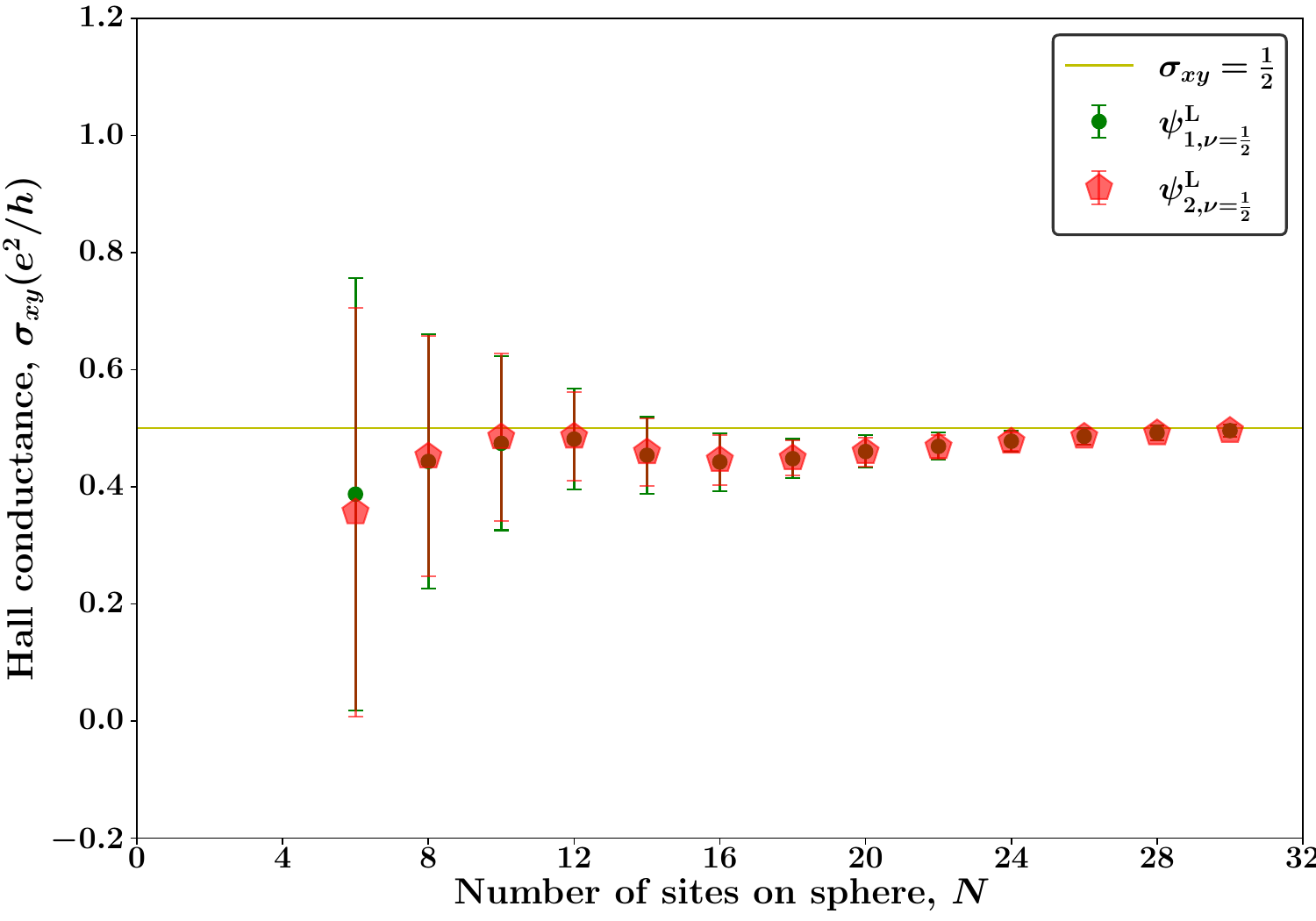}
        \caption{}
        \label{subfig:Hall_1_2_Laughlin}
    \end{subfigure}
    \end{minipage}
    \begin{minipage}[c]{0.28\textwidth}
    \begin{subfigure}{1\textwidth}
        \includegraphics[width=\linewidth]{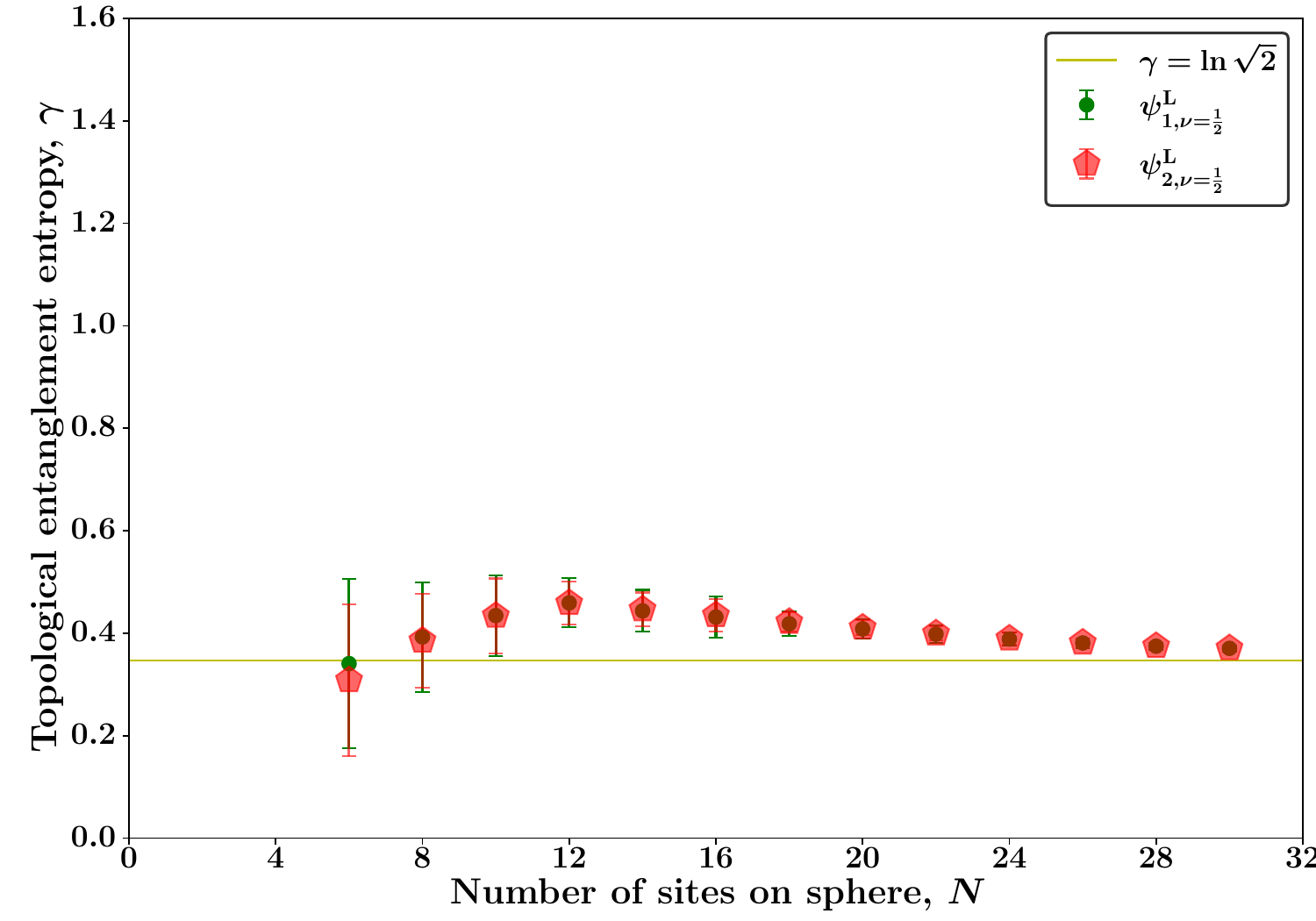}
        \caption{}
        \label{subfig:TEE_1_2_Laughlin}
    \end{subfigure}
    \end{minipage}
    \begin{minipage}[c]{0.28\textwidth}
    \begin{subfigure}{1\textwidth}
        \includegraphics[width=\linewidth]{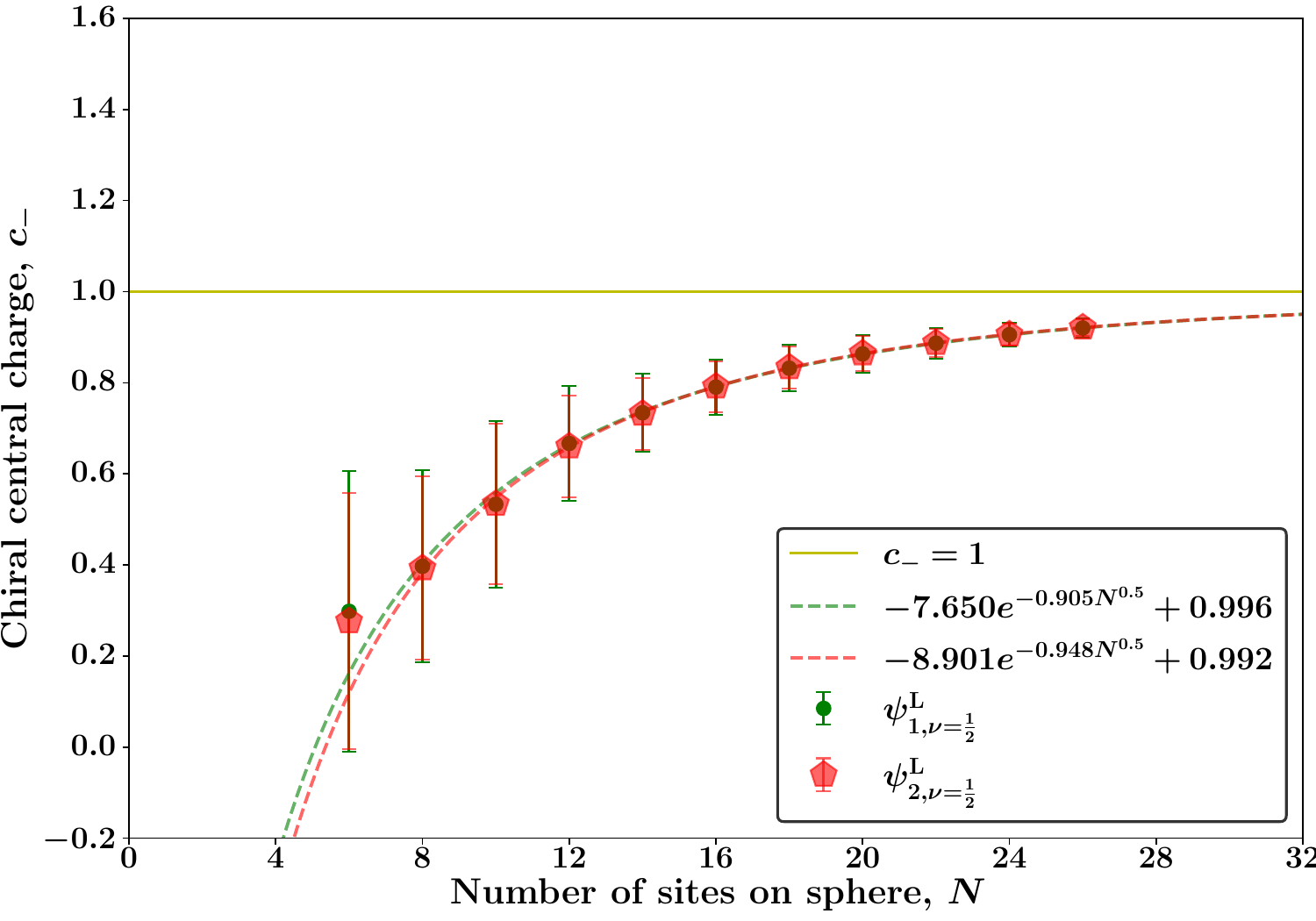}
        \caption{}
        \label{subfig:chiral_central_charge_1_2_Laughlin}
    \end{subfigure}
    \end{minipage}

    \vspace{0.25cm}

        \begin{minipage}[c]{0.12\textwidth}
    \begin{subfigure}[b]{1\textwidth}
        {\textbf{$\nu{=}1/3$ Laughlin}}
    \end{subfigure}
    \end{minipage}
    \begin{minipage}[c]{0.28\textwidth}
        \begin{subfigure}{1\textwidth}
        \includegraphics[width=\linewidth]{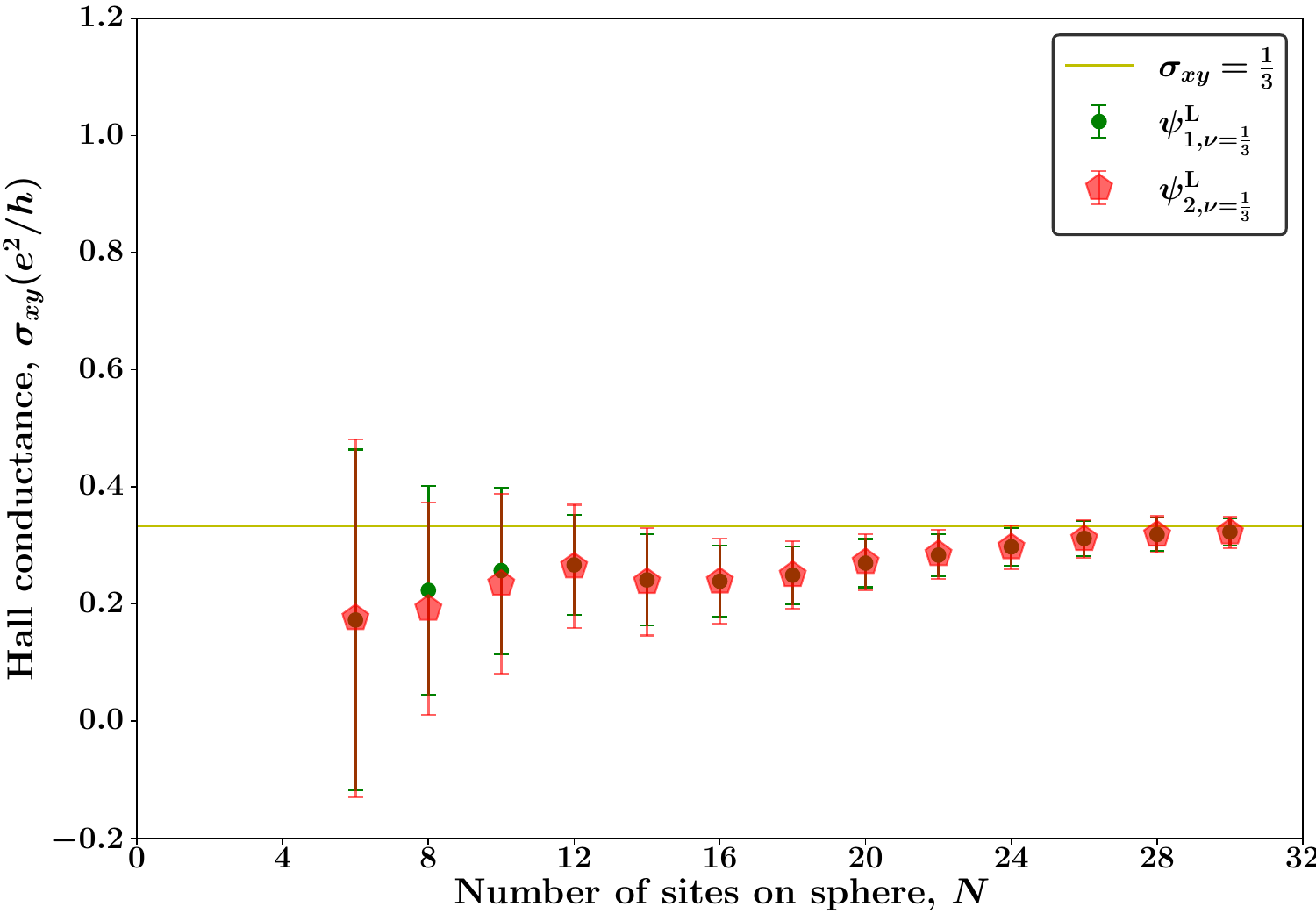}
        \caption{}
        \label{subfig:Hall_1_3_Laughlin}
    \end{subfigure}
    \end{minipage}
    \begin{minipage}[c]{0.28\textwidth}
    \begin{subfigure}{1\textwidth}
        \includegraphics[width=\linewidth]{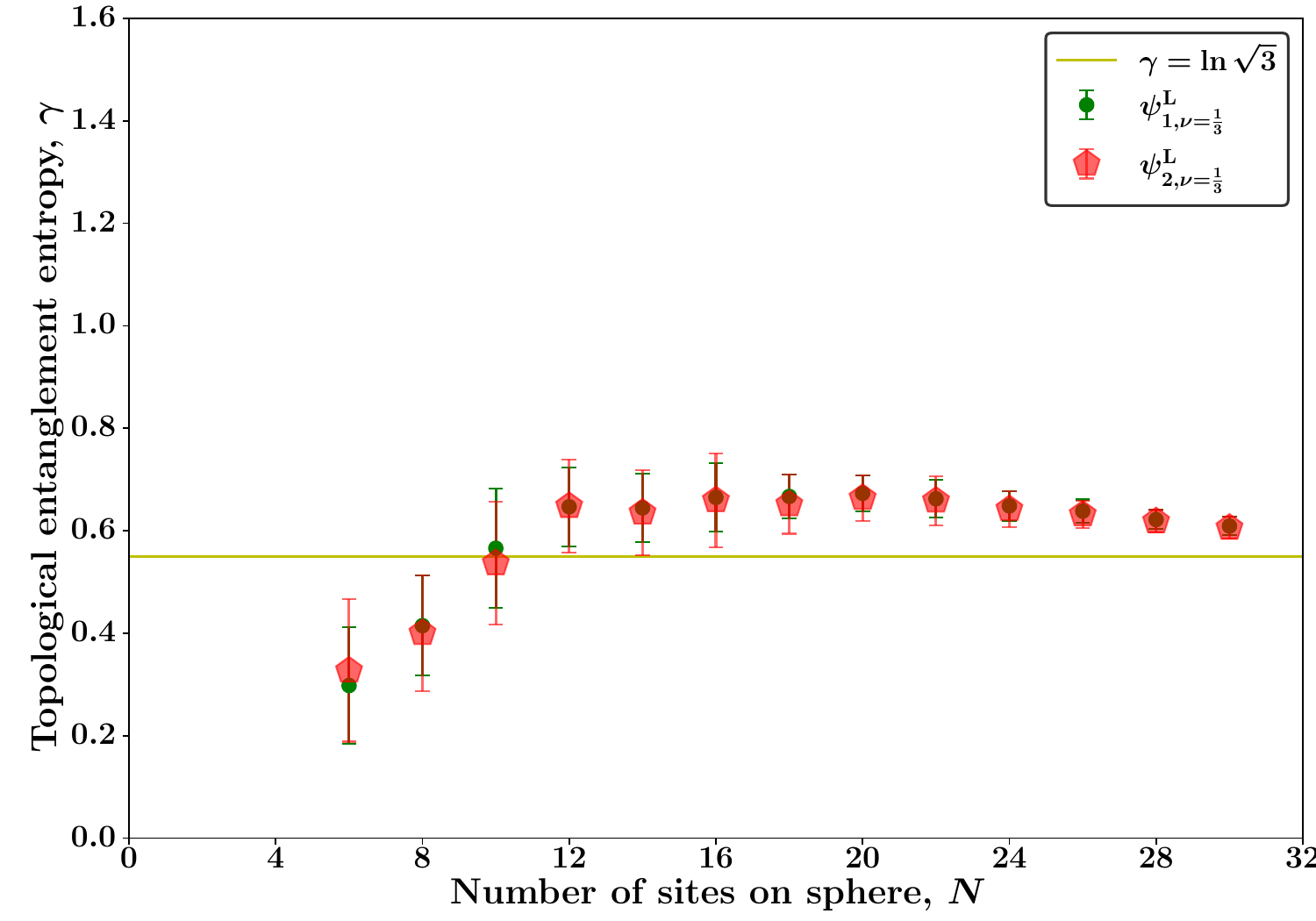}
        \caption{}
        \label{subfig:TEE_1_3_Laughlin}
    \end{subfigure}
    \end{minipage}
    \begin{minipage}[c]{0.28\textwidth}
    \begin{subfigure}{1\textwidth}
        \includegraphics[width=\linewidth]{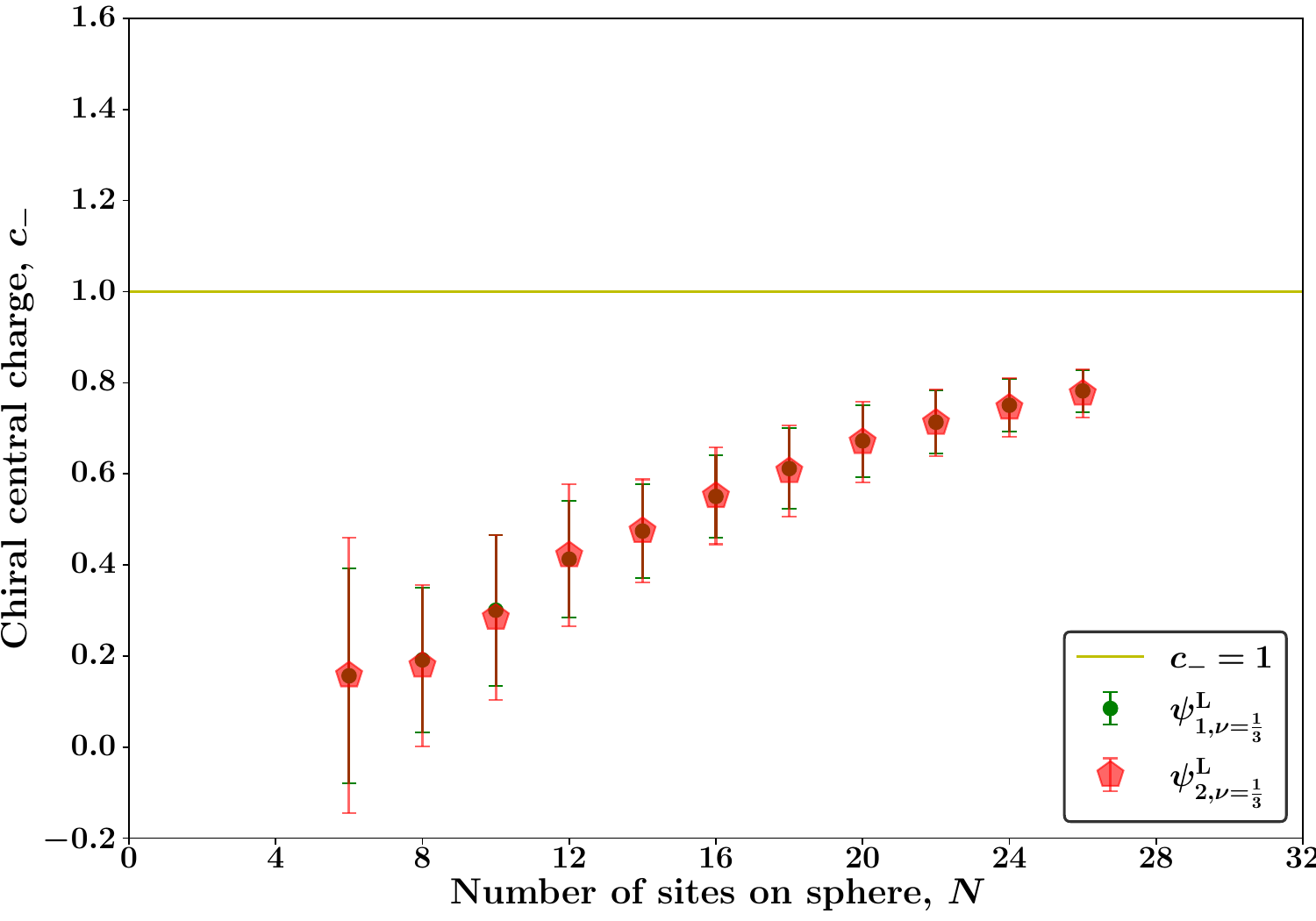}
        \caption{}
        \label{subfig:chiral_central_charge_1_3_Laughlin}
    \end{subfigure}
    \end{minipage}

    \vspace{0.25cm}

        \begin{minipage}[c]{0.12\textwidth}
    \begin{subfigure}[b]{1\textwidth}
        {\textbf{$\nu{=}1/4$ Laughlin}}
    \end{subfigure}
    \end{minipage}
    \begin{minipage}[c]{0.28\textwidth}
        \begin{subfigure}{1\textwidth}
        \includegraphics[width=\linewidth]{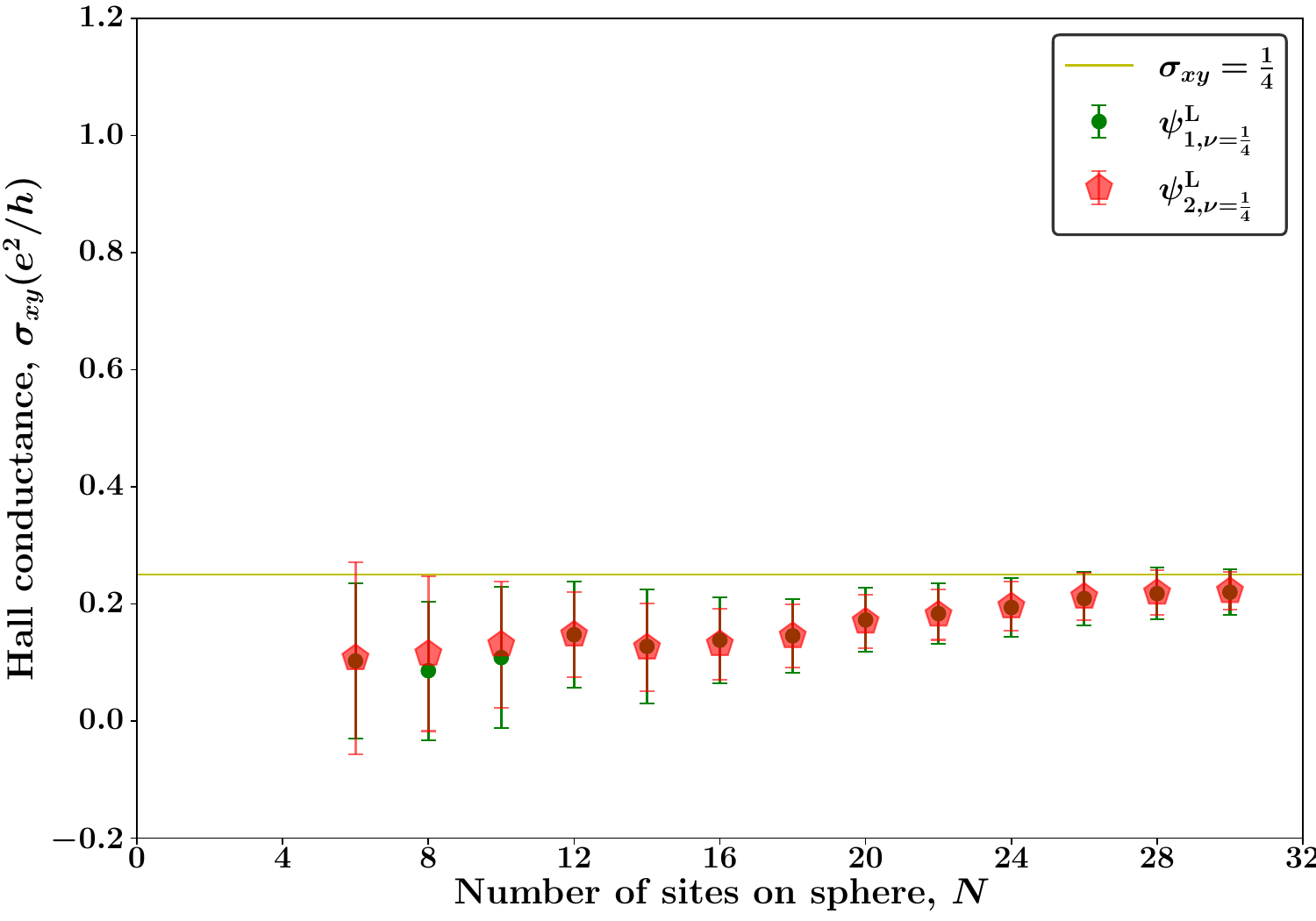}
        \caption{}
        \label{subfig:Hall_1_4_Laughlin}
    \end{subfigure}
    \end{minipage}
    \begin{minipage}[c]{0.28\textwidth}
    \begin{subfigure}{1\textwidth}
        \includegraphics[width=\linewidth]{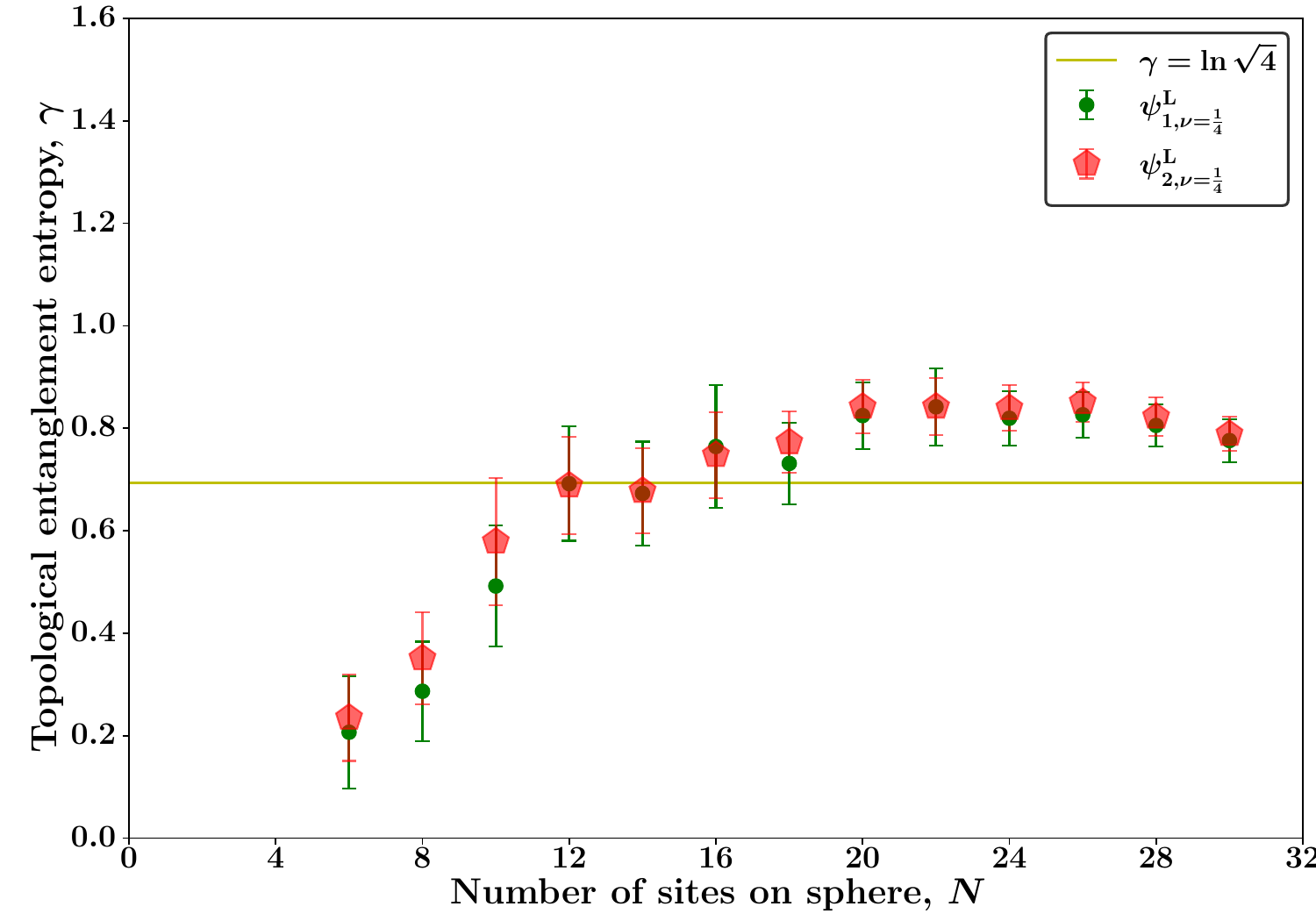}
        \caption{}
        \label{subfig:TEE_1_4_Laughlin}
    \end{subfigure}
    \end{minipage}
    \begin{minipage}[c]{0.28\textwidth}
    \begin{subfigure}{1\textwidth}
        \includegraphics[width=\linewidth]{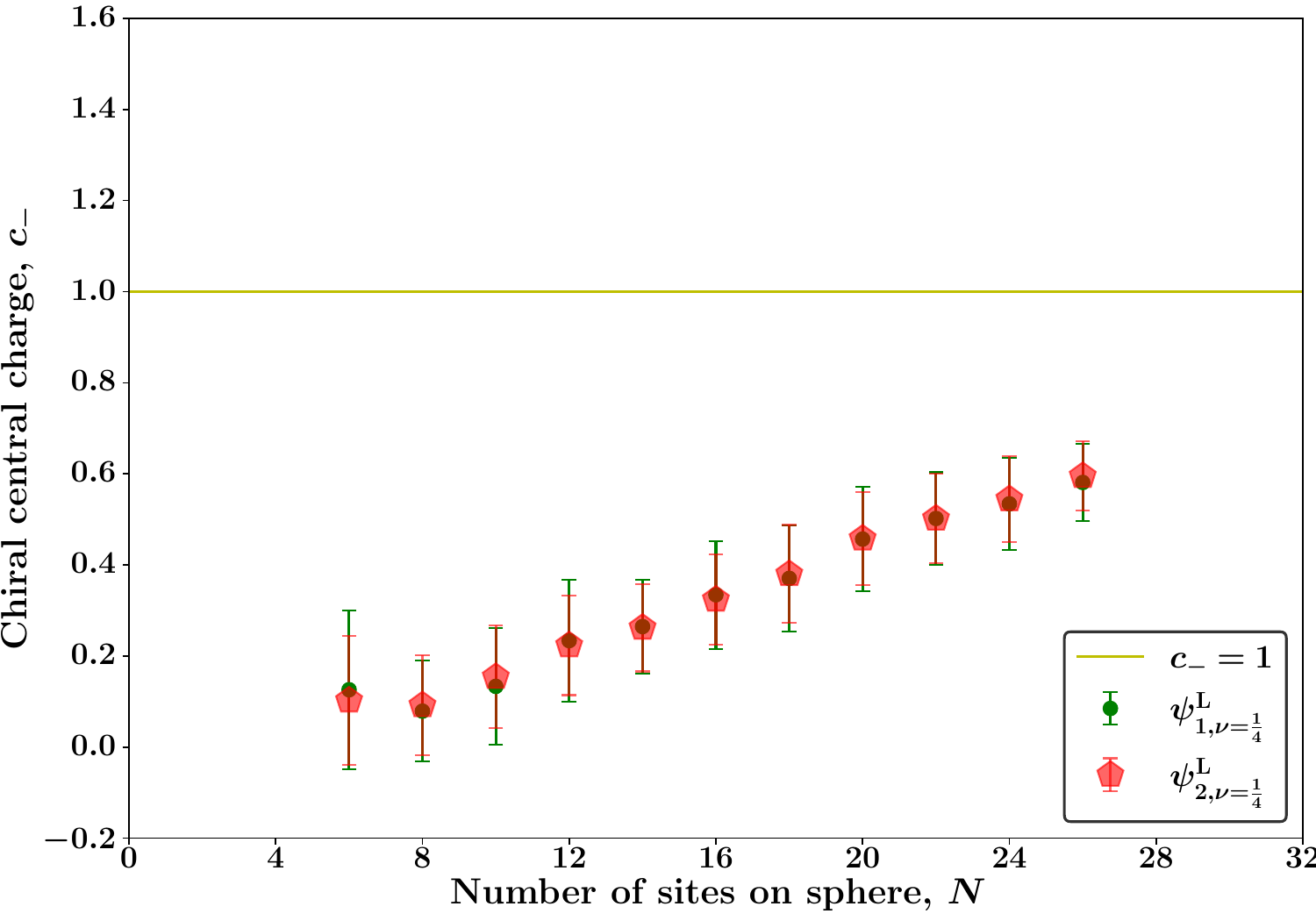}
        \caption{}
        \label{subfig:chiral_central_charge_1_4_Laughlin}
    \end{subfigure}
    \end{minipage}
    \caption{
      Mean values of Hall conductance, TEE, and chiral central charge have been plotted with error bars against the number of sites $N$ on the sphere. Along rows 1, 2, 3, and 4 are plotted the topological quantities for the IQH state at $\nu{=}1$ and the Laughlin state $1/2$, $1/3$, and $1/4$, respectively. Along the columns are plotted Hall conductance in column 1, TEE in column 2, and chiral central charge in column 3. The yellow line indicates the analytical value for the wavefunction in the $N{\rightarrow }\infty$ limit. The dashed line indicates the fit for the plot to a given wavefunction. The color of the fit and the color of the plot for the same wavefunction are kept the same.
    }
    \label{fig:all_topo_invariants_Laughlin}
\end{figure*}

\begin{figure*}[t]
    \begin{subfigure}{0.12\textwidth}
        {\footnotesize\textbf{state}}
    \end{subfigure}
    \begin{subfigure}{0.28\textwidth}
        \centering
        {\footnotesize \textbf{Hall conductance, $\sigma_{xy}$}}
    \end{subfigure}
    \begin{subfigure}{0.28\textwidth}
        {\footnotesize\textbf{topological entanglement entropy, $\gamma$}}
    \end{subfigure}
    \begin{subfigure}{0.28\textwidth}
        {\footnotesize\textbf{chiral central charge, $c_{-}$}}
    \end{subfigure}

    \vspace{0.3cm}

        \begin{minipage}[c]{0.12\textwidth}
    \begin{subfigure}[b]{1\textwidth}
        {\textbf{$\nu{=}1$ Moore-Read}}
    \end{subfigure}
    \end{minipage}
    \begin{minipage}[c]{0.28\textwidth}
        \begin{subfigure}{1\textwidth}
        \includegraphics[width=\linewidth]{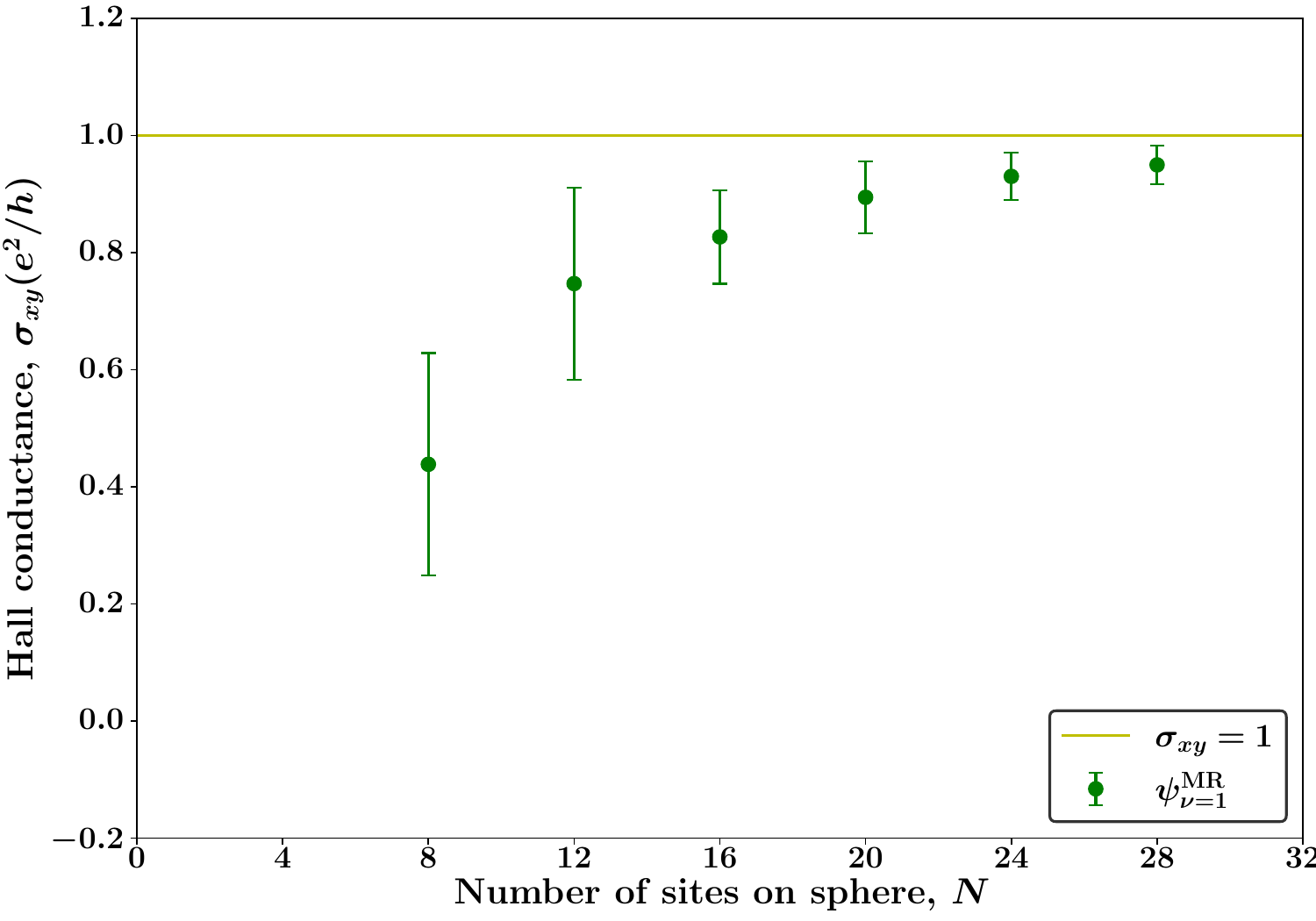}
        \caption{}
        \label{subfig:Hall_1_Moore_Read}
    \end{subfigure}
    \end{minipage}
    \begin{minipage}[c]{0.28\textwidth}
    \begin{subfigure}{1\textwidth}
        \includegraphics[width=\linewidth]{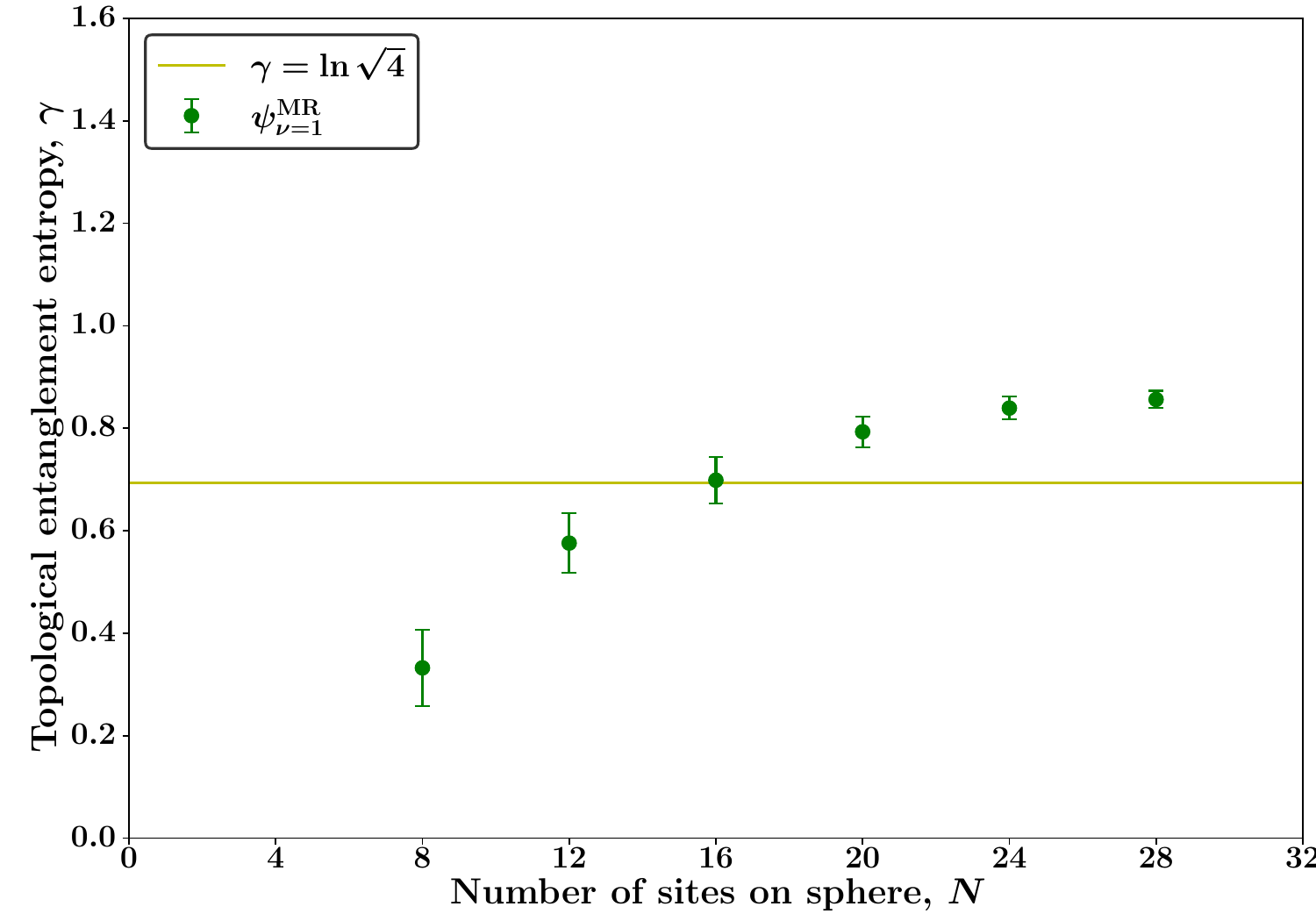}
        \caption{}
        \label{subfig:TEE_1_Moore_Read}
    \end{subfigure}
    \end{minipage}
    \begin{minipage}[c]{0.28\textwidth}
    \begin{subfigure}{1\textwidth}
        \includegraphics[width=\linewidth]{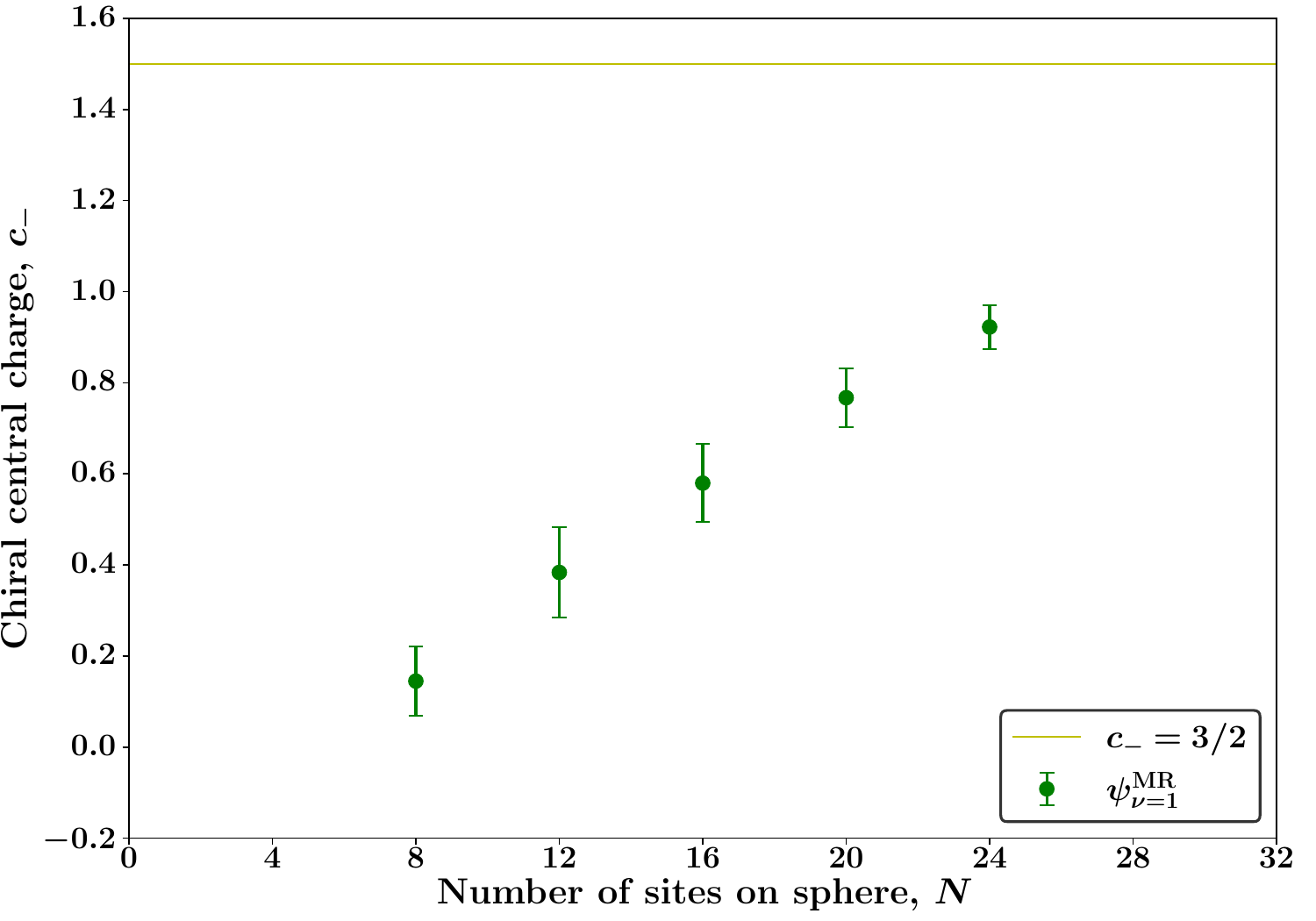}
        \caption{}
        \label{subfig:chiral_central_charge_1_Moore_Read}
    \end{subfigure}
    \end{minipage}

    \vspace{0.25cm}

        \begin{minipage}[c]{0.12\textwidth}
    \begin{subfigure}[b]{1\textwidth}
        {\textbf{$\nu{=}1/2$ Moore-Read}}
    \end{subfigure}
    \end{minipage}
    \begin{minipage}[c]{0.28\textwidth}
        \begin{subfigure}{1\textwidth}
        \includegraphics[width=\linewidth]{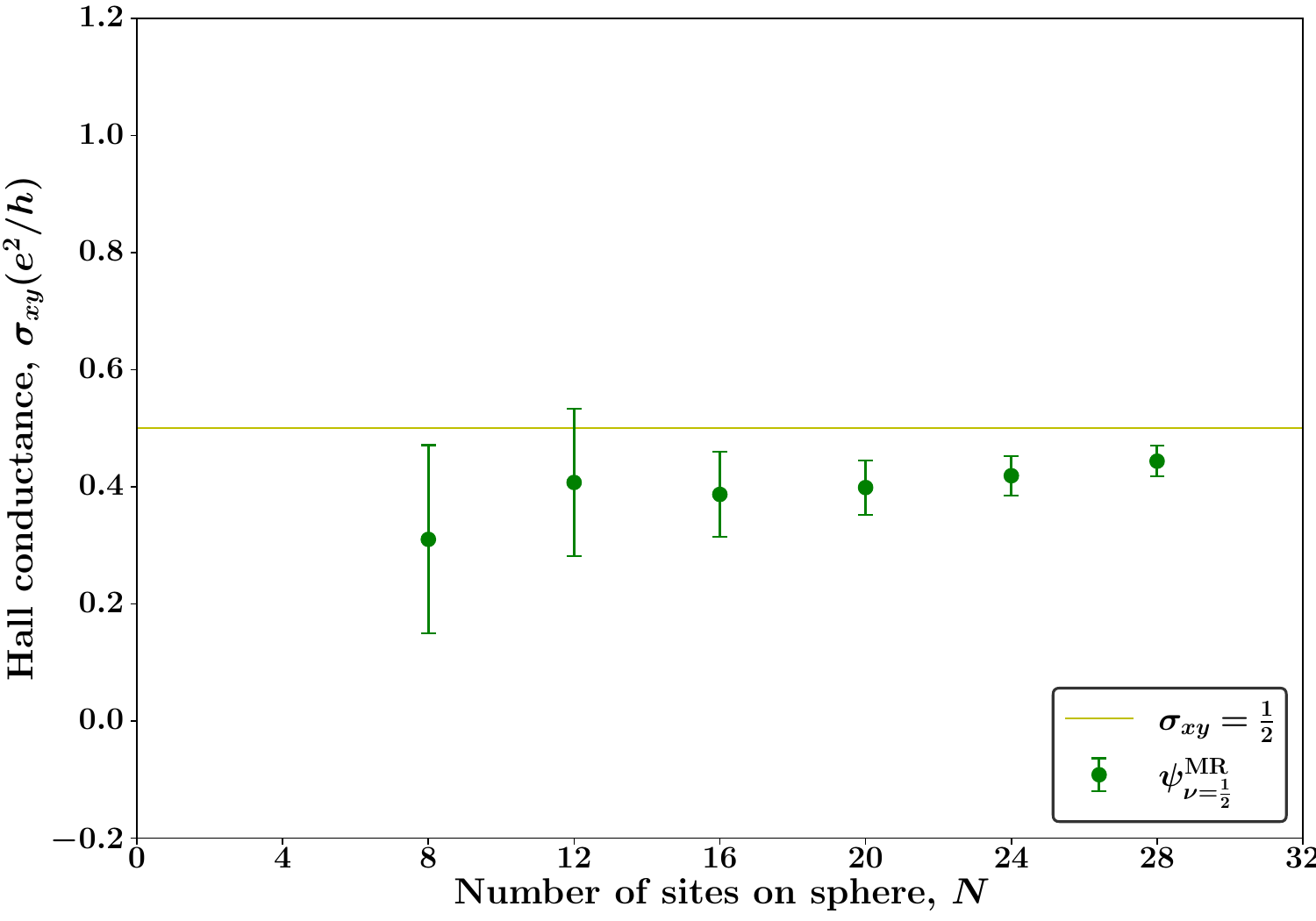}
        \caption{}
        \label{subfig:Hall_1_2_Moore_Read}
    \end{subfigure}
    \end{minipage}
    \begin{minipage}[c]{0.28\textwidth}
    \begin{subfigure}{1\textwidth}
        \includegraphics[width=\linewidth]{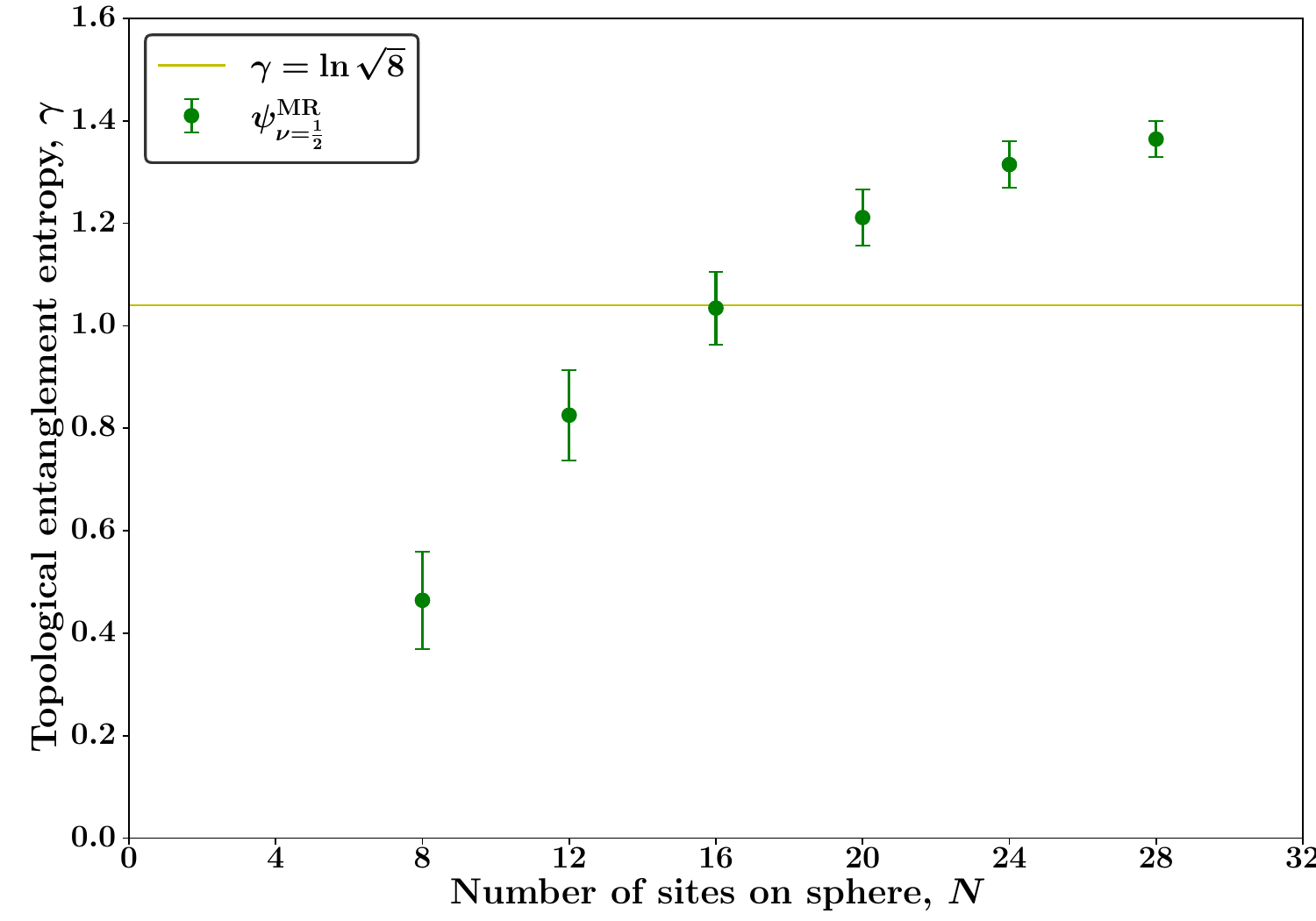}
        \caption{}
        \label{subfig:TEE_1_2_Moore_Read}
    \end{subfigure}
    \end{minipage}
    \begin{minipage}[c]{0.28\textwidth}
    \begin{subfigure}{1\textwidth}
        \includegraphics[width=\linewidth]{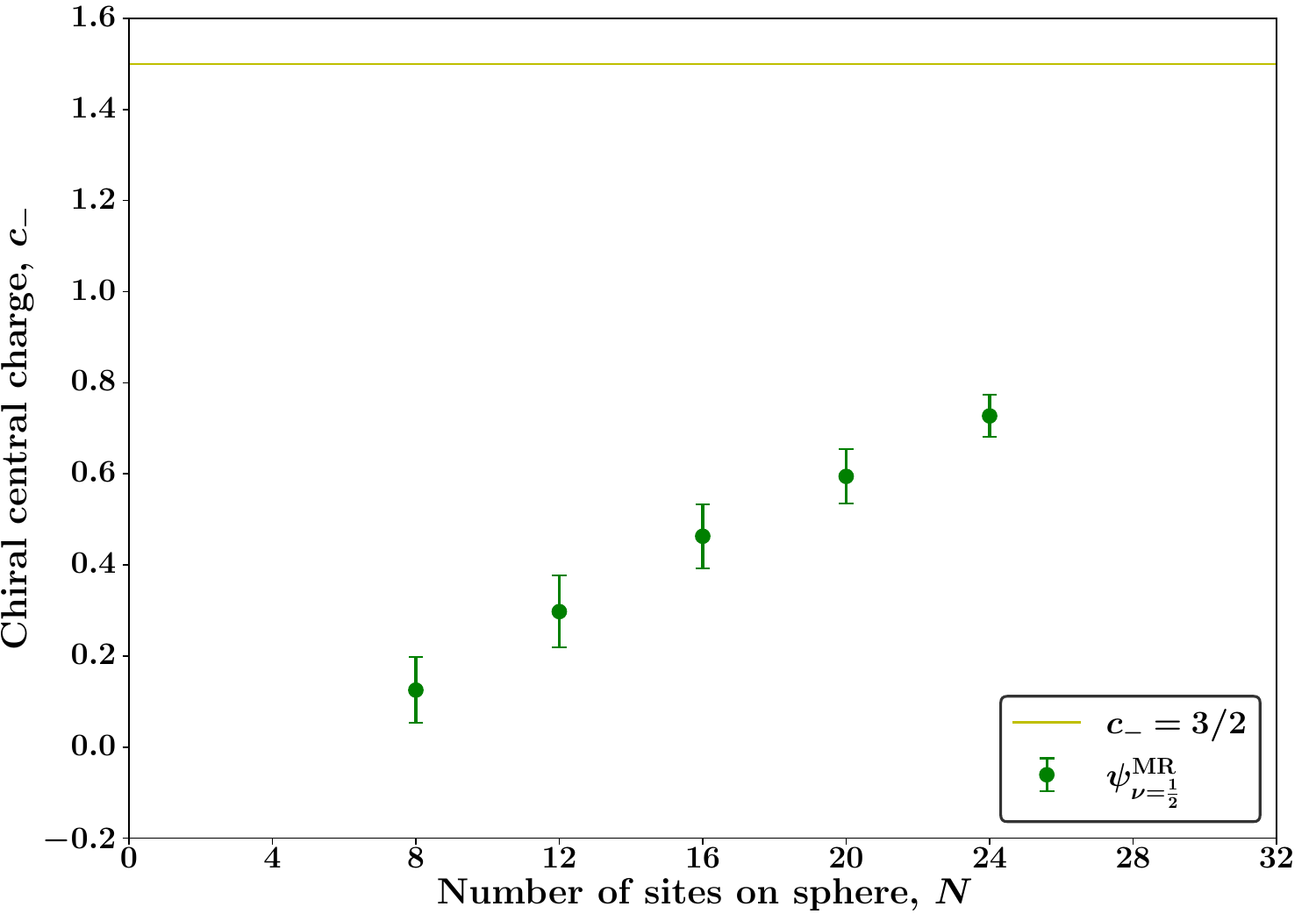}
        \caption{}
        \label{subfig:chiral_central_charge_1_2_Moore_Read}
    \end{subfigure}
    \end{minipage}

    \vspace{0.25cm}

        \begin{minipage}[c]{0.12\textwidth}
    \begin{subfigure}[b]{1\textwidth}
        {\textbf{$\nu{=}1/3$ Moore-Read}}
    \end{subfigure}
    \end{minipage}
    \begin{minipage}[c]{0.28\textwidth}
        \begin{subfigure}{1\textwidth}
        \includegraphics[width=\linewidth]{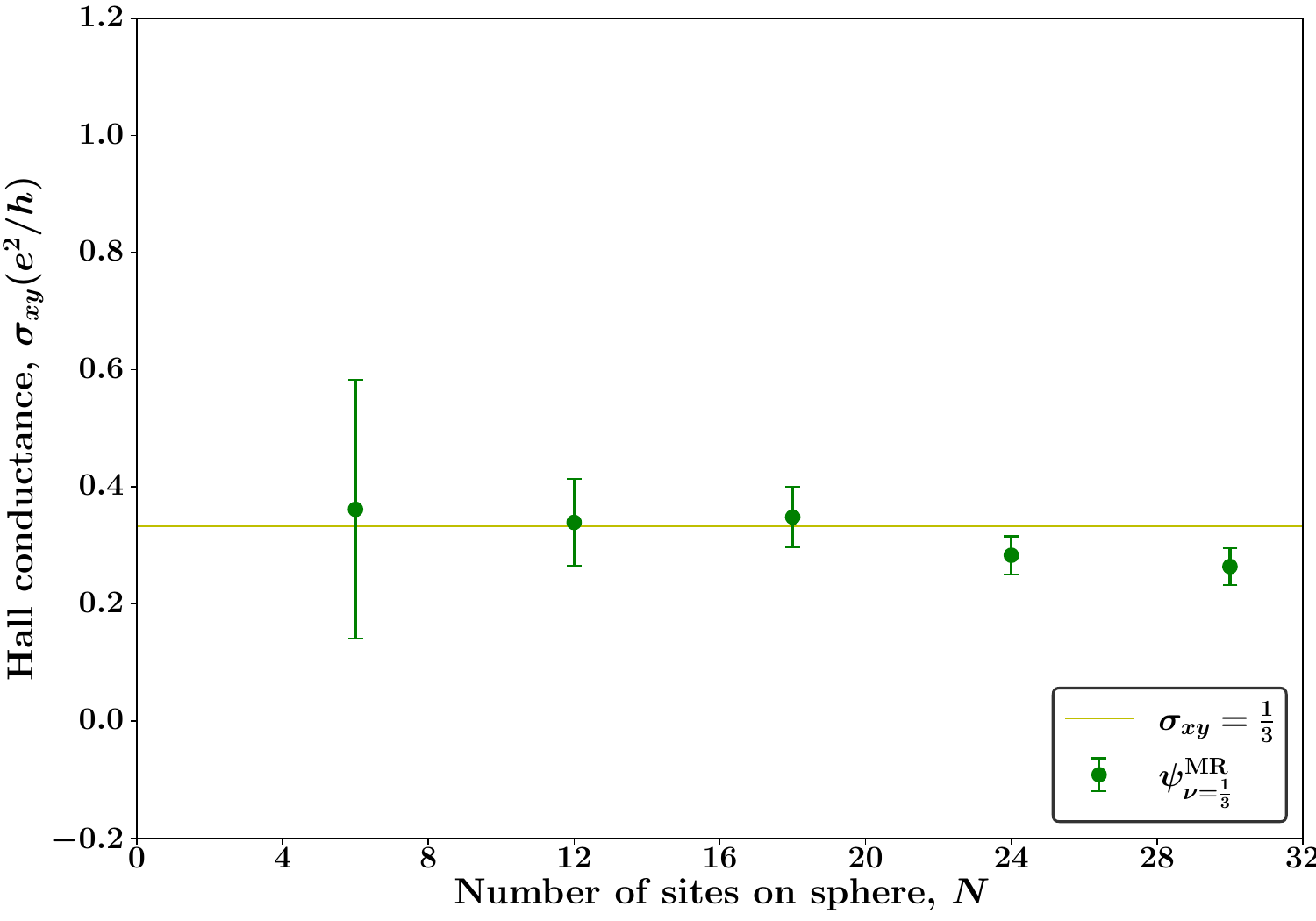}
        \caption{}
        \label{subfig:Hall_1_3_Moore_Read}
    \end{subfigure}
    \end{minipage}
    \begin{minipage}[c]{0.28\textwidth}
    \begin{subfigure}{1\textwidth}
        \includegraphics[width=\linewidth]{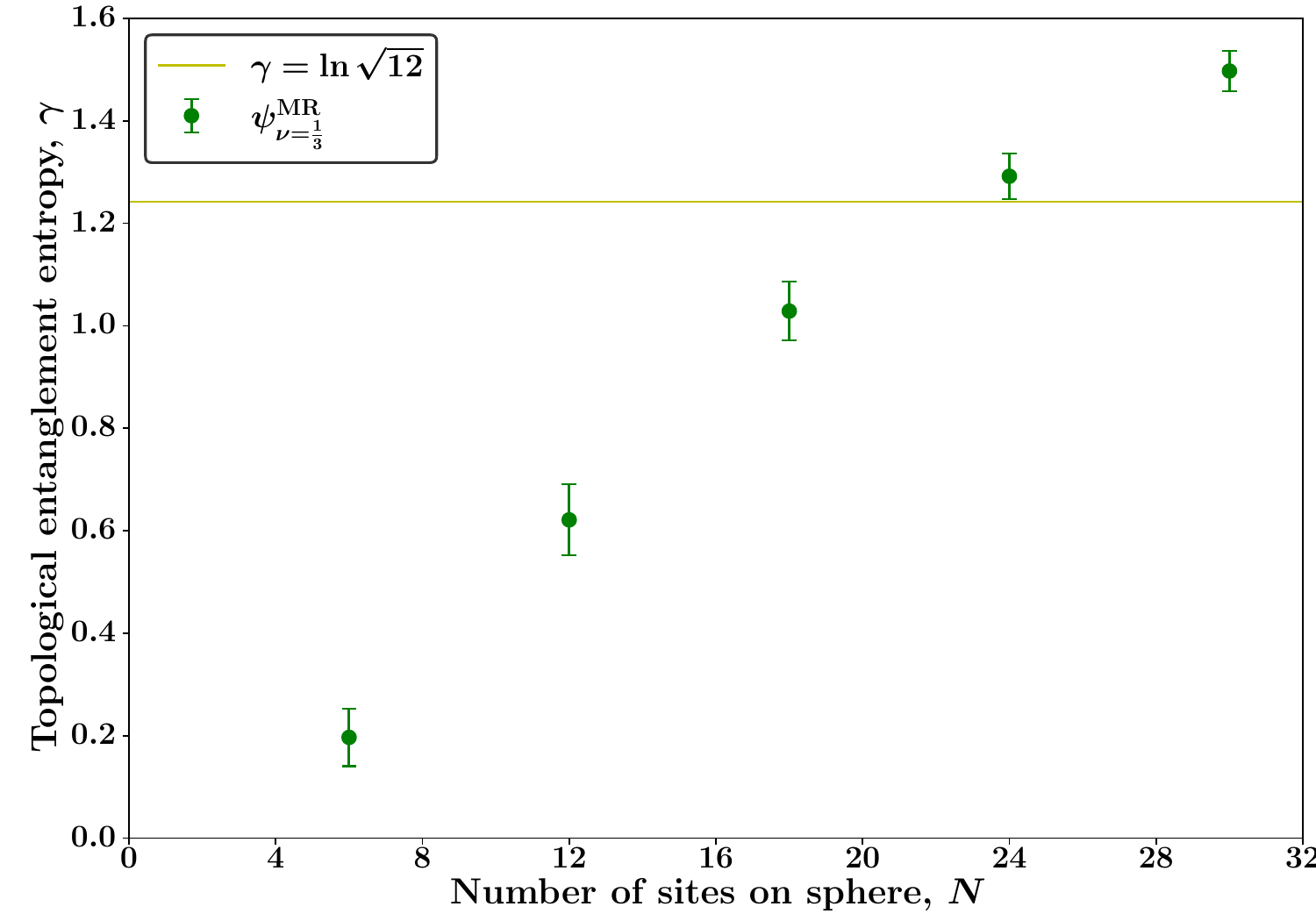}
        \caption{}
        \label{subfig:TEE_1_3_Moore_Read}
    \end{subfigure}
    \end{minipage}
    \begin{minipage}[c]{0.28\textwidth}
    \begin{subfigure}{1\textwidth}
        \includegraphics[width=\linewidth]{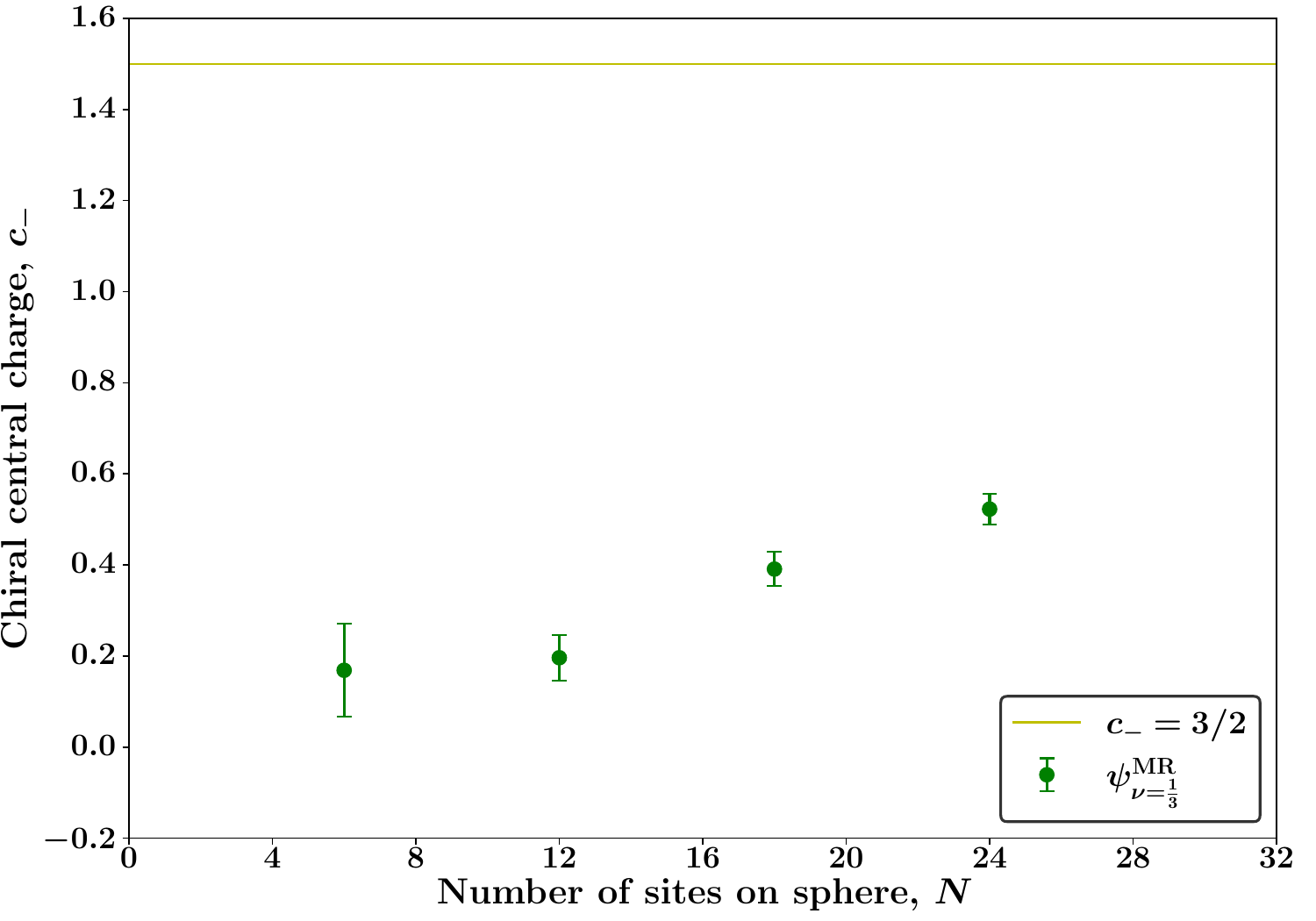}
        \caption{}
        \label{subfig:chiral_central_charge_1_3_Moore_Read}
    \end{subfigure}
    \end{minipage}
    \caption{
      Mean values of Hall conductance, TEE, and chiral central charge have been plotted with error bars against the number of sites $N$ on the sphere. Along rows 1, 2, and 3 are plotted the topological quantities for the Moore-Read state at $\nu{=}1$, $1/2$, and $1/3$, respectively. Along the columns are plotted Hall conductance in column 1, TEE in column 2, and chiral central charge in column 3. The yellow line indicates the analytical value for the wavefunction in the $N{\rightarrow }\infty$ limit. The dashed line indicates the fit for the plot to a given wavefunction. The color of the fit and the color of the plot for the same wavefunction are kept the same.
    }
    \label{fig:all_topo_invariants_Moore_Read}
\end{figure*}

The $N$ sites on the sphere are approximately uniformly distributed, with the coordinate of site $j$ being $\left(\sin(\theta_j)\cos(\phi_j),\sin(\theta_j)\sin(\phi_j),\cos(\theta_j)\right)$ with~\cite{Kim22}
\begin{equation}
    \label{eq: coordinate on sphere}
    (\theta_j,\phi_j)=\left(\cos^{-1}\left(1-\frac{2(j-1)}{N-1}\right),(j-1)\varphi\right),
\end{equation}
where $\varphi{=}2\pi(1{-}(\sqrt{5}{-}1)/2)$ is the golden angle and $j{\in}\{1,2,\dots,N\}$. To evaluate the topological quantities, the sphere is divided into four regions respecting tetrahedral symmetry. The distribution of the lattice sites on the sphere for $N{=}26$ is shown in Fig.~\ref{fig:sphere}.

The TEE, chiral central charge, and Hall conductance are calculated for various values of $N$. For each system size, we compute the average and standard deviation of these quantities over a certain number of uniform random rotations of the sphere. The averaging is done since the sites are not perfectly uniformly distributed, which results in variations in these topological quantities as the regions change.

For the Laughlin state, the constraint, $\sum_{i} s_i{=}0$, or equivalently, $\sum_i n_i{=}N/2$, restricts $N$ to be even. We compute the Hall conductance and TEE (chiral central charge) for all systems up to $N{=}30$ ($N{=}26$) for the IQH state at $\nu{=}1$ and Laughlin states at $\nu{=}1/2$, $1/3$, and $1/4$. For the Hall conductance and TEE (chiral central charge), for systems with $N{=}2$ to $24$ ($22$) sites, we do $12,800$ random rotations, while for $N{=}26, 28, 30$ ($N{=}24,26$), to keep the computation manageable, a smaller number of $1,280$ random rotations are done, over which the mean and standard deviations are calculated.

For the Moore-Read state, the constraint is $\sum_{i}n_i{=}\eta N/q$. In addition to this we need ${\rm Pf}_{n_i{=}n_j{=}1}\left[(u_i v_j {-} u_j v_i)^{{-}1}\right]$ of Eq.~\eqref{eq: Moore Read state wfn coefficient on sphere} to be non-zero, which happens only when $\eta N/q{=}{\text{even}}$. Choosing $\eta{=}1/2$ and $1$, we get $\sum_{i}n_i{=}N/(2q)$ and $N/q$, respectively. Setting $\eta{=}1/2$, $1$, and $1$ for the filling $\nu{=}1$, $1/2$, and $1/3$ which we consider, $N$ should be a multiple of $4$, $4$, and $6$, respectively. The Hall conductance and TEE (chiral central charge) for the $1$, $1/2$ Moore-Read are calculated for all systems with up to $N{=}28$ ($N{=}24$) sites. For the Hall conductance and TEE (chiral central charge), for systems with $N{=}4$ to $24$ ($20$) sites, we average over $12,800$ random rotations, while for $N{=}28$ ($N{=}24$), a smaller number of $1,280$ random rotations are averaged over. For $1/3$ Moore-Read, the Hall conductance and TEE (chiral central charge) are computed for system size up to $N{=}30$ ($N{=}24$) sites. For the Hall conductance and TEE (chiral central charge), systems with $N{=}6$ to $24$ ($18$) sites are averaged over $12,800$ random rotations, while for $N{=}30$ ($N{=}24$), averaging is done for of $1,280$ random rotations.

The main result of our work is shown in Figs.~\ref{fig:all_topo_invariants_Laughlin} and~\ref{fig:all_topo_invariants_Moore_Read}, where we plot the Hall conductance, TEE, and chiral central charge as a function of $N$ for the lattice model of the Laughlin and Moore-Read states, respectively. Next, we will analyze in detail each of these topological quantities.

\subsection{Hall conductance}
\label{ssec:result_hall}
\begin{table}[h]
\begin{tabular}{| c | c | c | c |}
\hline
state, $N_l$ & wavefunction & $\sigma_{xy}(N_l)$ & $\sigma_{xy}(N{\rightarrow}\infty)$ \\
\hline
 \multirow{8}{*}{\begin{tabular}{c} Laughlin \\ $N_l{=}30$ \end{tabular}} & $\psi_{1,\nu{=}1}^{\rm L}$ [Eq.~\eqref{eq: Laughlin on sphere coefficient 1}] & $0.998 {\pm} 0.001$ & \multirow{2}{*}{$1$}\\
 &  $\psi_{2,\nu{=}1}^{\rm L}$ [Eq.~\eqref{eq: Laughlin on sphere coefficient 2}] & $0.998{\pm}0.001$ & \\
 \cline{2-4}
 & $\psi_{1,\nu{=}1/2}^{\rm L}$ [Eq.~\eqref{eq: Laughlin on sphere coefficient 1}] & $0.496{\pm}0.01$ & \multirow{2}{*}{$1/2{=}0.5$}\\
 &  $\psi_{2,\nu{=}1/2}^{\rm L}$ [Eq.~\eqref{eq: Laughlin on sphere coefficient 2}] & $0.496{\pm}0.01$ & \\
 \cline{2-4}
 & $\psi_{1,\nu{=}1/3}^{\rm L}$ [Eq.~\eqref{eq: Laughlin on sphere coefficient 1}] & $0.32{\pm}0.03$ & \multirow{2}{*}{$1/3{=}0.33$}\\
 & $\psi_{2,\nu{=}1/3}^{\rm L}$ [Eq.~\eqref{eq: Laughlin on sphere coefficient 2}] & $0.32{\pm}0.02$ & \\
 \cline{2-4}
 & $\psi_{1,\nu{=}1/4}^{\rm L}$ [Eq.~\eqref{eq: Laughlin on sphere coefficient 1}] & $0.22{\pm}0.04$ & \multirow{2}{*}{$1/4{=}0.25$}\\
 & $\psi_{2,\nu{=}1/4}^{\rm L}$ [Eq.~\eqref{eq: Laughlin on sphere coefficient 2}] & $0.22{\pm}0.03$ & \\
 \hline
 \multirow{2}{*}{\begin{tabular}{c} Moore-Read \\ $N_l{=}28$ \end{tabular}} & $\psi^{\rm MR}_{\nu{=}1}$ [Eq.~\eqref{eq: Moore Read state wfn coefficient on sphere}] & $0.95{\pm}0.03$ & $1$\\
 \cline{2-4}
 &  $\psi^{\rm MR}_{\nu{=}1/2}$ [Eq.~\eqref{eq: Moore Read state wfn coefficient on sphere}] & $0.44{\pm}0.03$ & $1/2=0.5$\\
 \hline
 \begin{tabular}{c} Moore-Read \\ $N_l{=}30$ \end{tabular} & $\psi^{\rm MR}_{\nu{=}1/3}$ [Eq.~\eqref{eq: Moore Read state wfn coefficient on sphere}] & $0.26{\pm}0.03$ & $1/3{\sim}0.33$\\
\hline
\end{tabular}
\caption{\label{tab: Hall conductance} The Hall conductance, $\sigma_{xy}$, for the $\nu{=}1$ IQH state and the Laughlin states at $\nu{=}1/2$, $1/3$, and $1/4$ (obtained using two different wavefunctions) and the Moore-Read state at $\nu{=}1$, $1/2$, and $1/3$ for the largest system size ($N_l$) considered in this work. For comparison, in the last column, we show the theoretical expectation for $\sigma_{xy}$.}
\end{table}

The Hall conductance for the lattice model of the $\nu{=}1$ IQH state and the Laughlin states at $\nu{=}1/2$, $1/3$, and $1/4$ are plotted along column 1 of Fig.~\ref{fig:all_topo_invariants_Laughlin} and the $1$, $1/2$, and $1/3$ Moore-Read states are plotted along column 1 of Fig.~\ref{fig:all_topo_invariants_Moore_Read}. We find that the mean Hall conductance approaches its large $N$ limit with fluctuations that decrease with increasing $N$. The numerical values for the largest systems are tabulated in Table~\ref{tab: Hall conductance}. For the largest systems, the error is about $0.2\%$, $0.8\%$, $4\%$, and $12\%$ for the $\nu{=}1$ IQH state and the Laughlin states at $\nu{=}1/2$, $1/3$, and $1/4$, respectively, compared to the large $N$ limit. While for the Moore-Read at filling $1$, $1/2$ and $1/3$, the error is about $5\%$, 12\%, and $22\%$,, respectively. These results show that the Hall conductance converges to its expected value fastest for the $\nu{=}1$ IQH state, and slowest for the $1/3$ Moore-Read state.

\subsection{Topological entanglement entropy}
\label{ssec:result_TEE}
\begin{table}[h]
\begin{tabular}{| c | c | c | c |}
\hline
 state & wavefunction & $\gamma(N_{l})$ & $\gamma(N{\rightarrow}\infty)$ \\
\hline
 \multirow{8}{*}{\begin{tabular}{c} Laughlin \\ $N_l{=}30$ \end{tabular}} & $\psi_{1,\nu=1}^{\rm L}$ [Eq.~\eqref{eq: Laughlin on sphere coefficient 1}] & $0.007{\pm}0.001$ & \multirow{2}{*}{$\ln(\sqrt{1}){=}0$}\\
 & $\psi_{2,\nu=1}^{\rm L}$ [Eq.~\eqref{eq: Laughlin on sphere coefficient 2}] & $0.007{\pm}0.001$ & \\
 \cline{2-4}
 & $\psi_{1,\nu=1/2}^{\rm L}$ [Eq.~\eqref{eq: Laughlin on sphere coefficient 1}] & $0.370{\pm}0.006$ & \multirow{2}{*}{$\ln(\sqrt{2}){\sim }0.3466$}\\
 & $\psi_{2,\nu=1/2}^{\rm L}$ [Eq.~\eqref{eq: Laughlin on sphere coefficient 2}] & $0.371{\pm}0.005$ & \\
 \cline{2-4}
 & $\psi_{1,\nu=1/3}^{\rm L}$ [Eq.~\eqref{eq: Laughlin on sphere coefficient 1}] & $0.61{\pm}0.02$ & \multirow{2}{*}{$\ln(\sqrt{3}){\sim}$0.5493}\\
 & $\psi_{2,\nu=1/3}^{\rm L}$ [Eq.~\eqref{eq: Laughlin on sphere coefficient 2}] & $0.61{\pm}0.02$ & \\
 \cline{2-4}
 & $\psi_{1,\nu=1/4}^{\rm L}$ [Eq.~\eqref{eq: Laughlin on sphere coefficient 1}] & $0.78{\pm}0.04$ & \multirow{2}{*}{$\ln(\sqrt{4}){\sim}$0.6931}\\
 & $\psi_{2,\nu=1/4}^{\rm L}$ [Eq.~\eqref{eq: Laughlin on sphere coefficient 2}] & $0.79{\pm}0.03$ & \\
 \hline
 \multirow{2}{*}{\begin{tabular}{c} Moore-Read \\ $N_l{=}28$ \end{tabular}} & $\psi^{\rm MR}_{\nu=1}$ [Eq.~\eqref{eq: Moore Read state wfn coefficient on sphere}] & $0.86{\pm}0.02$ & $\ln(\sqrt{4}){\sim} 0.6931$\\
 \cline{2-4}
 & $\psi^{\rm MR}_{\nu=1/2}$ [Eq.~\eqref{eq: Moore Read state wfn coefficient on sphere}] & $1.36{\pm}0.04$ & $\ln(\sqrt{8}){\sim} 1.039$\\
\hline
 \begin{tabular}{c} Moore-Read \\ $N_l{=}30$ \end{tabular} & $\psi^{\rm MR}_{\nu=1/3}$ [Eq.~\eqref{eq: Moore Read state wfn coefficient on sphere}] & $1.50{\pm}0.04$ & $\ln(\sqrt{12}){\sim} 1.242$\\
\hline
\end{tabular}
\caption{\label{tab: TEE} The TEE, $\gamma$ for the $\nu{=}1$ IQH state and the Laughlin states at $\nu{=}1/2$, $1/3$, and $1/4$ (obtained using two different wavefunctions) and the Moore-Read state at $\nu{=}1$, $1/2$, and $1/3$ for the largest system size ($N_{l}$) considered in this work. For comparison, in the last column, we show the theoretical expectation for $\gamma$.}
\end{table}

The TEE for the lattice versions of the IQH state at $\nu{=}1$ and Laughlin state at $\nu{=}1/2$, $1/3$, and $1/4$ are plotted along column 2 of Fig.~\ref{fig:all_topo_invariants_Laughlin}. For the Moore-Read state, the TEE are plotted along column 2 of Fig.~\ref{fig:all_topo_invariants_Moore_Read} at $\nu{=}1$, $1/2$, and $1/3$. The TEE for the IQH and Laughlin states appear to converge to their expected values with increasing $N$, but for the Moore-Read states, the TEE continues to increase with $N$, up to the system sizes accessible. As with the Laughlin states, we expect the TEE to show a downturn (at a value of $N$ beyond the system sizes considered here) and then converge to the expected value as $N$ becomes large. 

Following Ref.~\cite{Kim22}, we fit the TEE for the $\nu{=}1$ IQH and $1/2$ Laughlin state to the form,
\begin{equation}
    \label{eq:fit_TEE}
    \gamma(N)=aN^{-k}+f,
\end{equation}
where $a$, $k$, $f$ are the fitting parameters. The fitting is done for $N{=}12$ to $30$. For the $\nu{=}1$ IQH state, the wavefunction of Eq.~\eqref{eq: Laughlin on sphere coefficient 1} gives $a{=}11.4 {\pm} 0.3$ and $f{=}{-}0.006 {\pm} 0.001$ for $k{=}2.0$, while for Eq.~\eqref{eq: Laughlin on sphere coefficient 2}, we find $a{=}11.5 {\pm} 0.4$ and $f{=}{-}0.006 {\pm} 0.001$ for $k{=}2.0$. The thermodynamic extrapolated value of the fit, $\gamma(\infty){=}{-}0.006$, is close to the analytical value, which is $\gamma{=}0$ since the non-interacting IQH state has no topological order.
For the Laughlin state at filling $1/2$ with the wavefunction of Eq.~\eqref{eq: Laughlin on sphere coefficient 1}, we obtain $a{=}0.89 {\pm} 0.02$ and $f{=}0.207 {\pm} 0.004$ for $k{=}0.5$, and $a{=}2.05 {\pm} 0.08$ and $f{=}0.302 {\pm} 0.003$ for $k{=}1$. Similarly, for Eq.~\eqref{eq: Laughlin on sphere coefficient 2} we find $a{=}0.93 {\pm} 0.03$ and $f{=}0.200 {\pm} 0.005$ for $k{=}0.5$, and for $k{=}1$, we get $a{=}2.1 {\pm} 0.1$ and $f{=}0.299 {\pm} 0.004$. This corroborates the fact that the two Laughlin wavefunctions of Eqs.~\eqref{eq: Laughlin on sphere coefficient 1} and~\eqref{eq: Laughlin on sphere coefficient 2} yield similar results for topological quantum numbers. Nevertheless, the thermodynamic extrapolated value of $\gamma{\approx}0.3$ for $k{=}1$ is still far away from the expected value of $\gamma{=}0.347$. We note that Ref.~\cite{Kim22}, carried out the fit to the TEE for the $1/2$ Laughlin state of Eq.~\eqref{eq: Laughlin on sphere coefficient 1}, and the fitted parameters they reported were $f{=}0.318{\pm} 0.005$ for $k{=}0.5$, and $f{=}0.217{\pm}0.006$ for $k{=}1$. Our analysis suggests that their reported values for $k{=}0.5$ and for $k{=}1$ got interchanged. Owing to the lack of sufficient systems post the downturn, we have not attempted to fit the data of other FQH states to the functional form of Eq.~\eqref{eq:fit_TEE}.

The numerical values for the largest systems are tabulated in Table~\ref{tab: TEE}. For the largest systems, the absolute error for $\nu{=}1$ IQH state is $0.007$ compared to its thermodynamic value, $\gamma{=}0$. For the Laughlin state at filling $1/2$, $1/3$, and $1/4$, the percentage error relative to its thermodynamic limit is about $7\%$, $11\%$, and $14\%$, respectively, while for Moore-Read state at filling $1$, $1/2$, and $1/3$, this error is about $24\%$, $31\%$, and $21\%$, respectively. As with the Hall conductance data, the convergence to the expected value is the fastest for the $\nu{=}1$ IQH state and slowest for the $1/3$ Moore-Read state.

\subsection{Chiral central charge}
\label{ssec:result_chiral}

\begin{table}[h]
\begin{tabular}{| c | c | c | c |}
\hline
 state & wavefunction & $c_{-}(N_{l})$ & $c_{-}(N{\rightarrow}\infty)$ \\
\hline
 \multirow{8}{*}{\begin{tabular}{c} Laughlin \\ $N_l{=}26$ \end{tabular}} & $\psi_{1,\nu=1}^{\rm L}$ [Eq.~\eqref{eq: Laughlin on sphere coefficient 1}] & $0.985{\pm}0.005$ & \multirow{8}{*}{$1$}\\
 &  $\psi_{2,\nu=1}^{\rm L}$ [Eq.~\eqref{eq: Laughlin on sphere coefficient 2}] & $0.984{\pm}0.005$ & \\
 \cline{2-3}
 & $\psi_{1,\nu=1/2}^{\rm L}$ [Eq.~\eqref{eq: Laughlin on sphere coefficient 1}] & $0.92{\pm}0.02$ & \\
 &  $\psi_{2,\nu=1/2}^{\rm L}$ [Eq.~\eqref{eq: Laughlin on sphere coefficient 2}] & $0.92{\pm}0.02$ & \\
 \cline{2-3}
 & $\psi_{1,\nu=1/3}^{\rm L}$ [Eq.~\eqref{eq: Laughlin on sphere coefficient 1}] & $0.78{\pm}0.05$ & \\
 & $\psi_{2,\nu=1/3}^{\rm L}$ [Eq.~\eqref{eq: Laughlin on sphere coefficient 2}] & $0.78{\pm}0.05$ & \\
 \cline{2-3}
 & $\psi_{1,\nu=1/4}^{\rm L}$ [Eq.~\eqref{eq: Laughlin on sphere coefficient 1}] & $0.58{\pm}0.08$ & \\
 & $\psi_{2,\nu=1/4}^{\rm L}$ [Eq.~\eqref{eq: Laughlin on sphere coefficient 2}] & $0.59{\pm}0.07$ & \\
 \hline
 \multirow{3}{*}{\begin{tabular}{c} Moore-Read \\ $N_l{=}24$ \end{tabular}} & $\psi^{\rm MR}_{\nu=1}$ [Eq.~\eqref{eq: Moore Read state wfn coefficient on sphere}] & $0.92{\pm}0.05$ & \multirow{3}{*}{$3/2{=}1.5$}\\
 \cline{2-3}
 & $\psi^{\rm MR}_{\nu=1/2}$ [Eq.~\eqref{eq: Moore Read state wfn coefficient on sphere}] & $0.73{\pm}0.05$ & \\
 \cline{2-3}
 & $\psi^{\rm MR}_{\nu=1/3}$ [Eq.~\eqref{eq: Moore Read state wfn coefficient on sphere}] & $0.52{\pm}0.03$ & \\
\hline
\end{tabular}
\caption{\label{tab: chiral_central_charges} The chiral central charge, $c_{-}$, for the $\nu{=}1$ IQH state and the Laughlin state at $\nu{=}1/2$, $1/3$ and $1/4$ (obtained using two different wavefunctions) and the Moore-Read state at $\nu{=}1$, $1/2$ and $1/3$ for the largest system size ($N_{l}$) considered in this work. For comparison, in the last column, we show the theoretical expectation for $c_{-}$.}
\end{table}

The chiral central charge for the lattice analogs of the $\nu{=}1$ IQH, and Laughlin states are plotted in column 3 of Fig.~\ref{fig:all_topo_invariants_Laughlin}, while those for the Moore-Read states are shown in column 3 of Fig.~\ref{fig:all_topo_invariants_Moore_Read}. The plots suggest that the chiral central charge could converge to its expected value with increasing $N$ for all the states. As with the Hall conductance and TEE, the convergence is rapid for the $\nu{=}1$ IQH and $1/2$ Laughlin state, but much slower for the other FQH states. 

We fit the chiral central charge for $N{=}12$ to $26$ for the $\nu{=}1$ IQH and $1/2$ Laughlin state to the following form proposed in Ref.~\cite{Kim22}
\begin{equation}
    \label{eq:fit_chiral}
    c_{-}(N)=ae^{-k\sqrt{x}}+f,
\end{equation}
where $a$, $k$, $f$ are fitting parameters. For $\nu{=}1$ IQH state with the wavefunction of Eq.~\eqref{eq: Laughlin on sphere coefficient 1}, we obtain the best fit for chiral central charge data at $k{=}0.81$ with $a{=}{-}1.95{\pm} 0.2$ and $f{=}1.018{\pm}0.004$ yielding an extrapolated value of $c_{-}(N{\rightarrow}\infty){=}1.018 {\pm} 0.004$. For Eq.~\eqref{eq: Laughlin on sphere coefficient 2}, the best fit is found at $k{=}0.81$ with $a{=}{-}1.98{\pm}0.2$ and $f{=}1.02 {\pm} 0.004$, resulting in $c_{-}(N{\rightarrow}\infty){=}1.02{\pm}0.004$. These values are close to the theoretically expected value of $c_{-}{=}1$.
For the $1/2$ Laughlin state of wavefunction of Eq.~\eqref{eq: Laughlin on sphere coefficient 1}, we get the best fit for chiral central charge data is found at $k{=}0.9048$ with $a{=}{-}7.65{\pm} 0.05$ and $f{=}0.9962{\pm}0.0007$ yielding an extrapolated value of $c_{-}(N{\rightarrow}\infty){=}0.9962 {\pm} 0.0007$. For the wavefunction of Eq.~\eqref{eq: Laughlin on sphere coefficient 2}, the best fit is obtained at $k{=}0.948$ with $a{=}{-}8.90{\pm}0.05$ and $f{=}0.9920 {\pm} 0.0007$, resulting in $c_{-}(N{\rightarrow}\infty){=}0.9920{\pm}0.0007$. The values are close to the theoretical result, $c_{-}{=}1$. We note that these results on the chiral central charge are fully consistent with those given in Ref.~\cite{Kim22}, which were obtained by using the wavefunction given in Eq.~\eqref{eq: Laughlin on sphere coefficient 1}. As with the TEE, owing to the lack of large systems, we have not attempted to do the fitting for the other FQH states. 

The numerical values for the largest systems are tabulated in Table~\ref{tab: chiral_central_charges}. The error for the largest systems, compared to the thermodynamic expectation, is about $1.6\%$, $8\%$, $22\%$, and  $42\%$ for $\nu{=}1$ IQH and $1/2$, $1/3$, $1/4$ Lauglin states, respectively. For Moore-Read state, this error is about $39\%$, $51\%$, and $65\%$ at filling $1$, $1/2$, and $1/3$, respectively. As with TEE and Hall conductance, the convergence to the $N{\to}\infty$ limit is the fastest for the $\nu{=}1$ IQH state, and slowest for the $1/3$ Moore-Read state.

We reiterate that the topological quantities for the Laughlin states obtained with the wavefunction given in Eqs.~\eqref{eq: Laughlin on sphere coefficient 1} and~\eqref{eq: Laughlin on sphere coefficient 2} are very close to each other, suggesting, as indicated before, that the extra factor $\prod_{1{\leq}j{<}k{\leq}N} (v_j u_k{-}v_k u_j)^{{-}2\alpha (n_j {+} n_k)}$ has little effect on the topological quantities even at finite $N$.

\section{Summary and conclusions}
\label{sec: summary}
We computed topological quantities, namely, the Hall conductance, TEE, and chiral central charge, for the lattice model of the $\nu{=}1$ IQH, and several Laughlin and Moore-Read states, and compared them to theoretical expectations. For the $\nu{=}1$ IQH and Laughlin states, as the filling decreases from $1$ to $1/4$, the convergence of the topological quantities to their thermodynamic values is slower. The convergence is even slower for the $1$, $1/2$, and $1/3$ Moore-Read states, with the convergence being the slowest for the $1/3$ Moore-Read state. This behavior can be understood in terms of the correlation length and nature of the underlying states. The correlation length increases as the filling decreases for the Abelian Laughlin states, while the non-Abelian Moore–Read state has an even larger correlation length~\cite{Estienne15} than the Laughlin state at the same filling [For states within a single class, such as Laughlin, the larger the topological entanglement entropy (or equivalently, its total quantum dimension~\cite{Kitaev06a, Levin06}), the larger is the correlation length. However, it is trickier to compare states across different classes, for example, the Abelian $1/4$ Laughlin and the non-Abelian $1$ Moore-Read state both have $\gamma{=}\ln(2)$, but the former has a larger correlation length than the latter.]. Consequently, stronger finite-size effects are observed at lower fillings, and to reliably recover the topological data at low fillings, larger system sizes than those considered in this work are required. However, this is challenging to do since the modular techniques involve evaluating the singular value decomposition of a matrix whose size grows factorially with $N$, thereby requiring significantly larger computing resources for larger systems. Additionally, we only considered ideal model wavefunctions, not exact eigenstates of an interacting Hamiltonian, such as the Coulomb Hamiltonian. We anticipate that extracting topological quantum numbers from these eigenstates will be even more challenging since fewer system sizes are accessible through exact diagonalization.  

We note that recently, it was shown that the modular commutator can be used to identify chiral topological order in certain spin models~\cite{Maity25}. The ground states in these spin models map to bosonic FQH states. R\'enyi generalizations~\cite{Gass25}, and other entanglement probes~\cite{Sheffer25, Sheffer25a} that have been developed recently, can also be applied to the states considered in this work to see how well they capture the topological order.

\begin{acknowledgments}
We acknowledge useful discussions with Bowen Shi. The work was made possible by financial support from the Science and Engineering Research Board (SERB) of the Department of Science and Technology (DST) via the Mathematical Research Impact Centric Support (MATRICS) Grant No. MTR/2023/000002. Computational portions of this research work were conducted using the Nandadevi and Kamet supercomputers maintained and supported by the Institute of Mathematical Sciences' High-Performance Computing Center. 
\end{acknowledgments}

\bibliography{refrences, biblio_fqhe}

\end{document}